\documentclass[12pt,preprint]{aastex}

\newcommand{\text}[1]{\rm #1}

\slugcomment{}

\shorttitle{Ray-Tracing Analysis for Gravitational Radiation in Core-Collapse 
Supernovae}
\shortauthors{Kotake et al.}

\begin{document}

\title{Ray-Tracing Analysis of Anisotropic Neutrino Radiation
 for Estimating Gravitational Waves in Core-Collapse Supernovae}

\author{Kei Kotake\altaffilmark{1}, Wakana Iwakami\altaffilmark{2}, 
Naofumi Ohnishi\altaffilmark{2}, and Shoichi Yamada\altaffilmark{3,4}}
\affil{$^1$Division of Theoretical Astronomy/Center for Computational Astrophysics, National Astronomical Observatory of Japan, 2-21-1, Osawa, Mitaka, Tokyo, 181-8588, Japan}
\email{kkotake@th.nao.ac.jp}
\affil{$^2$Department of Aerospace Engineering, Tohoku University,
6-6-01 Aramaki-Aza-Aoba, Aoba-ku, Sendai, 980-8579, Japan}
\affil{$^3$Science \& Engineering, Waseda University, 3-4-1 Okubo, Shinjuku,
Tokyo, 169-8555, Japan}
\affil{$^4$Advanced Research Institute for Science and Engineering, Waseda University, 3-4-1 Okubo, Shinjuku,Tokyo, 169-8555, Japan}
\begin{abstract}
We propose a ray-tracing method to estimate gravitational waves (GWs)
generated by anisotropic neutrino emission in supernova cores. 
To calculate the gravitational waveforms, we derive analytic formulae in a useful form, 
which are applicable also for three-dimensional computations.
 Pushed by evidence of slow rotation prior to core-collapse,
 we focus on asphericities in neutrino emission and matter motions outside the 
  protoneutron star. 
Based on the two-dimensional (2D) models,
 which mimic SASI-aided neutrino heating explosions, 
 we compute the 
 neutrino anisotropies via the ray-tracing method in a post-processing manner 
and calculate the resulting waveforms. 
For simplicity, neutrino absorption and emission 
by free nucleons, dominant processes outside the PNSs, are only taken into account, 
 while the neutrino scattering and the velocity-dependent terms in the transport 
equations are neglected. With these computations, 
it is found that the waveforms exhibit more variety 
in contrast to the ones previously estimated by the ray-by-ray analysis
(e.g., Kotake et al. (2007)).
In addition to a positively growing feature,
which was predicted to determine the total wave amplitudes predominantly, 
the waveforms are shown to exhibit 
large negative growth for some epochs during the growth of SASI. 
These features are found to stem 
from the excess of neutrino emission in lateral directions,
 which can be precisely captured by the ray-tracing 
 calculation.
Reflecting the nature of SASI which grows chaotically with time,
 there is little systematic dependence of the input neutrino luminosities on the 
maximum wave amplitudes.
 Due to the negative contributions and 
 the neutrino absorptions appropriately taken into account by the ray-tracing method, 
 the wave amplitudes become more than one-order-of magnitude smaller 
than the previous estimation, thus 
making their detections very hard for a galactic source.
On the other hand, it is pointed out that 
the GW spectrum from matter motions have its peak 
 near  $\sim 100$ Hz, reflecting the SASI-induced matter overturns of $O(10)$ ms.
 Such a feature could be characteristic for the SASI-induced supernova explosions.
 The proposed ray-tracing method
 will be useful for the GW prediction in the 
first generation of 3D core-collapse supernova simulations 
that do not solve the angle-dependent neutrino transport equations 
as part of the numerical evolution.
\end{abstract}
\keywords{supernovae: collapse --- gravitational waves --- 
neutrinos --- hydrodynamics}

\clearpage

\section{Introduction}
No longer gravitational-wave astronomy is a fantasy.
In fact, 
gravitational wave detectors, such as
 LIGO \citep{firstligo,firstligonew},
VIGRO\footnote{http://www.ego-gw.it/},
GEO600\footnote{http://geo600.aei.mpg.de/},
TAMA300 \citep{tama,tamanew},
and AIGO\footnote{http://www.gravity.uwa.edu.au/}
with their 
international network of the observatories, 
are beginning to 
take data at sensitivities where astrophysical
events are predicted 
(see, e.g., \citet{hough} for a recent review).  
For the detectors, core-collapse supernovae
especially in our Galaxy, have been proposed as one of the most plausible sources 
of gravitational waves (see, for example, \citet{kotake_rev,ott_rev} 
for recent reviews). Since the gravitational waves (plus neutrinos) are the only tool 
that gives us the live information of the central engine of core-collapse supernovae,
 the detection is important not only for the direct confirmation of general relativity 
but also for disclosing the supernova physics itself. 

Traditionally, most of the theoretical predictions of gravitational waves (GWs)
  have focused on the bounce signals
(e.g., \citet{zweg,kotakegw,shibaseki,ott,obergaulinger,cerda,dimmelprl,simon,dimmel_prd} and references 
therein). However recent stellar evolution calculations 
suggest that rapid rotation assumed in most of the previous studies is not canonical 
for the progenitors with neutron star formation
\citep{heger05}.
To explain the observed rotation periods of radio pulsars, 
the rotation periods of the iron core before collapse are estimated to be larger than 
$\sim$ 100 sec 
\citep{ott_birth}. In such a slowly rotating case, the detection of the 
bounce signals becomes very hard even by the next-generation laser interferometers 
for a Galactic supernova (e.g., \citet{kotakegwmag}).

Besides the rapid rotation of the core, 
convective matter motions and anisotropic neutrino emission in the much 
later postbounce phase are expected to be the primary GW sources 
with comparable amplitudes to the bounce signals. 
Thus far, various physical ingredients for producing asphericities
 and the resulting GWs in the postbounce phase, have been studied such as the roles 
of pre-collapse density inhomogeneities \citep{burohey,muyan97,fryersingle,fryer04},
 moderate rotation of the iron core \citep{mueller}, 
g-mode oscillations of protoneutron stars (PNSs) 
\citep{ott_new},  and SASI \citep{kotake_gw_sasi,marek_gw}.

Among them, we focused on the GWs originated from the 
asphericities produced by the standing accretion shock instability (SASI) \citep{kotake_gw_sasi}. Here SASI, becoming 
very popular in current supernova researches, 
is a uni- and bipolar sloshing of the stalled supernova shock 
with pulsational strong expansion and contraction (see, e.g., \citet{blondin_03,scheck_04,ohnishi_1,fogli07,blondin07a,iwakami08,iwakami08_2} and references therein). 
Based on the two-dimensional (2D) simulations, which demonstrate the neutrino-driven explosions aided by SASI using the light-bulb scheme
 (see \citet{jankamueller96,ohnishi_1} for details), it was pointed out 
that the GW amplitudes from anisotropic 
neutrino emission increase almost monotonically with time, which are 
dominant over the ones from matter motions, and that such signals
 may be visible to next-generation detectors for a Galactic source.
More recently, \citet{marek_gw} analyzed the GW emission based on their long-term 2D 
 Boltzmann (ray-by-ray) simulations, which seem very close to produce 
the SASI-aided neutrino-driven explosions
 \citep{marek}. They also found that the GWs from neutrinos 
with continuously growing amplitudes (but with the different sign of the 
amplitudes in \citet{kotake_gw_sasi}), are dominant over the ones from matter 
motions. They proposed that the third-generation class detectors 
such as the Einstein Telescope are required for detecting the GW signals 
with a good signal-to-noise ratio.

Except for the acoustic mechanism \citep{ott_new} and 
the magnetohydrodynamic mechanism
 (e.g., \citet{yamasawa,kotakemhd,ard,shibata_mhd,burrows_07,takiwaki_mhd} and 
references therein), both of which produce strong mass-quadrupole GWs,  
all the studies in the postbounce phase mentioned above, rely basically 
on the conventional 
neutrino-heating mechanism \citep{bethe}. They agree 
that the GW amplitudes from anisotropic neutrino emission are
 dominant over the ones from mass motions.
 This means that accurate estimation of the neutrino anisotropy 
is indispensable for understanding the gravitational radiation from core-collapse 
supernovae, 
which requires to estimate precisely the directional dependence 
of the neutrino intensity emitted from the central cores.
In the previous simulations with the light-bulb approximations 
\citep{muyan97,kotake_gw_sasi}, there was no way but to 
 estimate the angle-dependent neutrino luminosities, assuming that neutrinos are emitted 
purely radially in each lateral bin of the computational polar grid.
 This ray-by-ray treatment cannot capture the neutrino emission in the lateral 
 directions entirely.
The FLD(flux-limited-diffusion) schemes (either multi-energy \citep{walder} 
or single-energy FLD \citep{burohey,fryersingle,fryer04}) tend to smooth 
out the local and global neutrino anisotropies due to their diffusion characters.
   Any ray-by-ray transport schemes 
\citep{buras,marek}, albeit coupled to the Boltzmann transport and thus being one 
of the most sophisticated treatment at present,
 replaces the 2D neutrino transport with the 1D transport problem along radial rays in every lateral bin, leading to the overestimation of
the directional dependence of the neutrino anisotropies
(see discussions in \citet{marek_gw}).  
More recently, fully 2D multi-angle Boltzmann 
transport simulations 
 become practicable \citep{ott_ray}, however too computationally expensive currently
to perform the simulations, 
 satisfying required number of the momentum-space angles in order to capture the neutrino anisotropies accurately (e.g., \citet{ott_rev}). 

These situations motivate us to propose a method to estimate the neutrino 
anisotropies, directly linked to the accurate estimation of 
 the neutrino-originated gravitational radiation. Pushed by the striking evidences 
that support the slow rotation of iron cores \citep{heger05,ott_birth}, 
we consider an idealized situation that 
the neutrino radiation field from the PNSs are isotropic. 
This means that no GWs are assumed to be generated inside due to the isotropy.
Then we focus on the asphericities outside the PNSs, which are produced by 
the growth of SASI.
 To this end, we utilize the 2D SASI models as in our previous paper 
\citep{kotake_gw_sasi}, in which the neutrino luminosities from the
PNSs are adjusted to trigger explosions aided by SASI \citep{ohnishi_1}.
 Since the regions outside the PNSs are basically optically thin to neutrinos,  
 we solve the transport equations by making use of 
the ray-tracing (long-characteristics) method in a post-processing manner.
For the sake of simplicity, we consider here only neutrino absorptions and emission 
by free nucleons, dominant processes outside the PNS, 
 while the neutrino scattering and the velocity-dependent terms in the transport 
equation are neglected.
 We then study how the obtained gravitational waveforms 
could change from the ones obtained in the ray-by-ray analysis 
(e.g., \citet{kotake_gw_sasi}) and discuss their implications.
 Although the presented scheme is designed to be 
 valid for the regions where neutrinos are thin to matter, 
 we hope it to be useful for the GW prediction in 
the first generation of 3D core-collapse supernova 
simulations that do not solve the transport equations as part of the numerical 
evolution. 

The plan of this paper is as follows. In section \ref{sec2}, 
first the formalism for calculating gravitational waveforms is described. 
Thereafter, we describe the ray-tracing method to calculate the direction-dependent 
neutrino luminosities. In Section \ref{sec3}, 
 we shall briefly summarize the information how to construct the 2D exploding models, 
 such as about the initial models and numerical methods.
The main results are shown in Section \ref{sec4}.  We summarize our results 
and discuss their implications in Section \ref{sec5}.  

\section{Computing the Gravitational Wave Signatures \label{sec2}}

\subsection{Formulae for Gravitational Waves from Anisotropic 
Neutrino Radiation \label{sec2.1}}
To compute the gravitational waveforms from anisotropic neutrino 
radiation, we follow the formalism pioneeringly proposed by \cite{epstein}.
 In the following, we present the formulae in a useful form, which is applicable 
 also for 3D computations.

\begin{figure}[hbt]
\epsscale{0.4}
\plotone{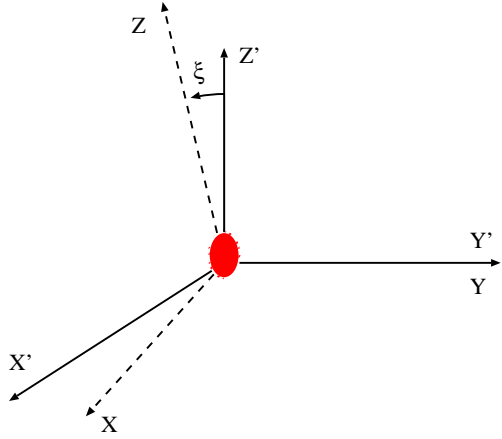}
\caption{Source coordinate system $(x',y',z')$ and observer
 coordinate system  $(x,y,z)$. The observer resides at the
 distant point on the $z$-axis. The viewing angle is denoted by $\xi$
 which is the angle between $z$ and $z'$ axis. The $z'$ axis coincides with the
 symmetry axis of the source, presumably the rotational axis. Central
 red region illustrates the anisotropic neutrino radiation from a core-collapse 
supernova.}
\label{f1}
\end{figure}

The two polarization states (of $+$ and $\times$ modes) of
 GWs from anisotropic neutrino radiation satisfying
the transverse-traceless conditions are given by \citet{muyan97} as follows,
\begin{equation}
h_{+} = \frac{2G}{c^4 R}\int_{0}^{t}  dt'
\int_{4 \pi} d\Omega' (1 + \cos \theta)~\cos2\phi~
\frac{dl_{\nu}(\Omega',t')}{d\Omega'},
\label{t_+}
\end{equation}
and 
\begin{equation} 
 h_{\times} =  \frac{2G}{c^4 R}\int_{0}^{t}  dt'
\int_{4 \pi} d\Omega' (1 + \cos \theta)~\sin2 \phi~
\frac{dl_{\nu}(\Omega',t')}{d\Omega'},
\label{t_cross}
\end{equation}
where $G$ is the gravitational constant, $c$ is the speed of light, $R$
is the distance of the source to the observer 
, $dl_{\nu}/d\Omega$
represents the direction-dependent neutrino luminosity emitted per unit
of solid angle into direction of $\Omega$.
Variables with dash such as $\Omega'$ represent the
quantities of the source coordinate system, while non-dashed ones
belong to the the observer coordinate system (for the
geometrical setup, see Figure \ref{f1}). For convenience,
 we assume that $y$ axis coincides with $y'$ axis and that 
the $z$-axis lies on the ($x',z'$) plane. 

\begin{figure}[h]
\epsscale{0.8}
\plotone{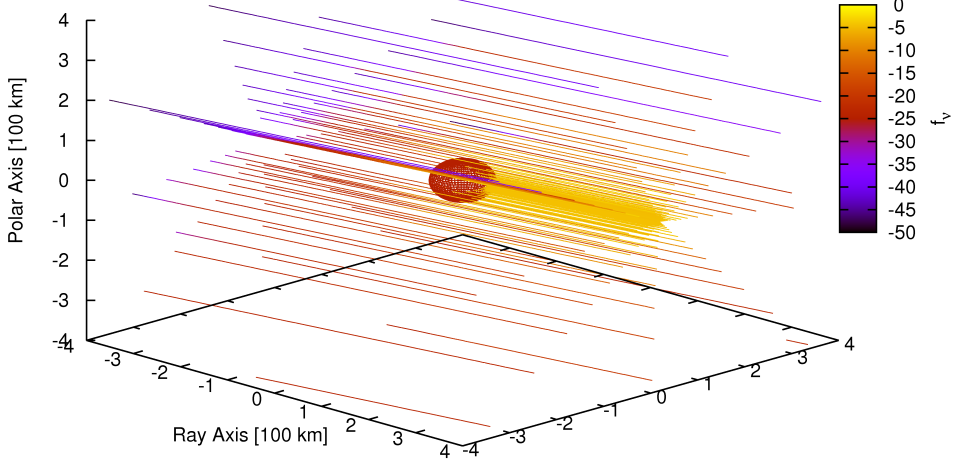}
\caption{An example of ray-tracing of neutrinos for estimating 
$dl_{\nu}({\bf\Omega})/d\Omega$ towards a given direction of ${\bf\Omega}$.
The central region colored by red represents the surface of the protoneutron star (PNS)
(located at 50 km in radius), which is the inner boundary of our computation.
 The color-scale on the rays shows the logarithmic values of $f_{\nu}$, 
the neutrino occupation 
 probability (for $\epsilon_{\nu} = 14$ MeV here), 
which is calculated by the line integral along each ray (equation (\ref{transfer})). 
For this snapshot taken from our 2D simulations, 
the higher values of 
 $f_{\nu}$ are seen to come just in front of the PNS (yellow on the rays), while
$f_{\nu}$ becomes smaller in the distant regions from the PNS.
Doing ray-tracing calculations along every direction, 
$dl_{\nu}({\bf\Omega})/d\Omega$ can be estimated through equation (\ref{final}).}
\label{ray_trace}
\end{figure}

Using the following relations between the two coordinates,
\begin{equation}
\sin \theta \cos \phi = \sin \theta' \cos \phi' \cos \xi -  \cos \theta' \sin \xi,
\end{equation}
\begin{equation}
\sin \theta \sin \phi = \sin \theta' \sin \phi',
\end{equation}
\begin{equation}
\cos \theta =  \sin \theta' \cos \phi' \sin \xi + \cos \theta' \cos \xi,
\end{equation}
 $\theta$ and $\phi$ in equations (\ref{t_+}, \ref{t_cross}) are
required to be expressed in terms of the angles $\theta',
~\phi'$ with respect to the source coordinate valuables,
and the viewing angle of $\xi$. After some algebra,  one can readily derive the 
 two modes of the GWs (equations (\ref{t_+}, \ref{t_cross})) as follows,
\begin{eqnarray}
h_{+} &=& \frac{2G}{c^4 R}\int_{0}^{t} dt'
\int_{4 \pi} d\Omega' 
(1 + \sin \theta' \cos \phi' \sin \xi + \cos \theta' \cos \xi) \times \nonumber \\ 
& & \frac{(\sin \theta' \cos\phi' \cos \xi - \cos \theta' \sin \xi)^2 - \sin^2 \theta' \sin^2 \phi'}{(\sin \theta' \cos\phi' \cos \xi - \cos \theta' \sin \xi)^2 + \sin^2 \theta' \sin^2 \phi'}
\frac{dl_{\nu}(\Omega',t')}{d\Omega'},
\label{t+}
\end{eqnarray}
and 
\begin{eqnarray}
 h_{\times} &=&  \frac{4G}{c^4 R}\int_{0}^{t} dt'
\int_{4 \pi} d\Omega'  (1 + \sin \theta' \cos \phi' \sin \xi + \cos \theta' \cos \xi) \times \nonumber \\ 
& & \frac{\sin \theta' \sin \phi'(\sin \theta' \cos\phi' \cos \xi - \cos \theta' \sin \xi)}
{(\sin \theta' \cos\phi' \cos \xi - \cos \theta' \sin \xi)^2 + \sin^2 \theta' \sin^2 \phi'}
\frac{dl_{\nu}(\Omega',t')}{d\Omega'},
\label{tcc}
\end{eqnarray}
 which will be useful in computing the GW signals also for 3D computations. 
It is noted that the sum of the squared
amplitudes $|h_{+}|^2 + |h_{\times}|^2$ is invariant under the rotation
about the $z$-axis. To maximize 
 the GW amplitudes in our 2D axisymmetric case, we assume that the observer 
is situated along the direction of the equatorial plane ($\xi = \pi/2$) 
as in \citet{kotake_gw_sasi}. Then the only non-vanishing component is, 
\begin{eqnarray}
h^{\rm e}_{+} \equiv h_{\nu} &=& \frac{2G}{c^4 R} \int_{0}^{t} dt'
\int_{4 \pi}~d\Omega' (1 + \sin \theta' \cos \phi')
\frac{\cos^2 \theta' - \sin^2 \theta' \sin^2 \phi'}
{\cos^2 \theta' + \sin^2 \theta' \sin^2 \phi'}
\frac{dl_{\nu}(\theta^{'},t^{'})}{d\Omega^{'}}\nonumber \\
&=&  \frac{4G}{c^4 R} \int_{0}^{t} dt^{'}
\int_{0}^{\pi}~d\theta'~\Phi(\theta')~\frac{dl_{\nu}(\theta',t')}
{d\Omega'}.
\label{tt}
\end{eqnarray}
Here the subscripts of ($^{e}$) and ($_\nu$) indicates that 
the observer is situated in the equatorial plane and 
that the GWs are originated from neutrinos.
$\Phi(\theta^{'})$ depends on the angle measured from the symmetry axis 
($\theta^{'}$) 
\begin{equation}
\Phi({\theta^{'}})=  \pi \sin \theta^{'} ( - 1 + 2 | \cos  \theta^{'}| ).
\label{graph1}
\end{equation} 
As given in Figure 1 of \citet{kotake_gw_sasi}, this function
 has positive values in the north polar cap for $0 \leq \theta' \leq 60^{\circ}$ and in 
the south polar cap for $120^{\circ} \leq \theta' \leq 180^{\circ}$, 
but becomes negative values between $60^{\circ} < \theta' < 120^{\circ}$. 
 In order to perform numerically the angular integration 
in equation (8) with high numerical accuracy, 
we recommend to perform it based on a Gaussian quadrature,
 because the function of $\Phi$ of equation 
(\ref{graph1}) is not smooth near poles and equator.
For estimating $h_{\nu}$ (equation (8)), 
 we are yet to determine the directional dependent 
neutrino luminosities of $dl_{\nu}/d\Omega$, which we will estimate by the ray-tracing 
method in the next section. 

Here it is noted that GWs generated by neutrinos are distinct from 
the ones from matter dynamics, because the former 
 has the {\it memory effect}, which means that the 
gravitational amplitude jumps from zero to a nonvanishing value and 
it keeps the non-vanishing value even after the energy 
source of gravitational waves disappeared (see \citet{braginsky} for details). 
 In equations (\ref{t_+},\ref{t_cross}), this nature can be directly seen as the 
time-integral. Other astrophysical emitters 
of GWs with memory have been elaborately studied for 
 gamma-ray bursts \citep{hiramatsu,suwa_murase}, Pop III stars \citep{suwa1,suwa2}, and 
 inspiralling compact objects \citep{favata} (see references therein).

As for the gravitational waves 
of the quadrupole radiation of mass motions, we employ the stress formula
 (see, e.g., equation (12) in \cite{mm}). In using the formula, we consider the 
self-gravity of matter in the accretion flow. In the following computations, we 
assume that the source is assumed to be located at our galactic
center ($R = 10~\rm{kpc}$) and also that the observer is 
situated in the direction of the equatorial plane.

\subsection{Ray-Tracing Calculations of Anisotropic Neutrino Luminosities  \label{sec2.2}}

Now we proceed to determine $dl_{\nu}/d\Omega$ in equation (8), 
the directional dependent neutrino luminosities through the ray-tracing method. 
 Note in the following equations that we change dashed variables to non-dashed ones 
for the sake of simplicity. 

In the ray-tracing approach, we consider transfer along the ray specified by 
a constant impact parameter $p$. The coordinate along $p$ is called $s$, satisfying
\begin{equation}
r = (p^2 + s^2)^{1/2},
\end{equation}
where $r$ is the radial coordinate. To estimate $dl_{\nu}({\bf\Omega})/d\Omega$ along
 a given direction of ${\bf\Omega}$, we draw rays of neutrinos as shown in Figure 
\ref{ray_trace}. As will be discussed soon later, in order to get the numerical 
convergence of $dl_{\nu}({\bf\Omega})/d\Omega$, 
 we need to set 45,000 rays for each direction, which consists of 
$500 \times 90$ rays, where the former is 
for the impact parameters covering from the inner- ($p_{\rm in}$ = 50 km) to the outer- 
boundary ($p_{\rm out} $ = 2000 km)
 of the computational domain and the latter is for covering 
the circumference (e.g., $2\pi$) of the concentric circles on the plane perpendicular to the rays (see Figure \ref{ray_trace}). In the axisymmetric case here, we 
perform the ray-tracing calculations 60 times to cover the entire sphere, 
which is the number of the mesh points for the polar direction (e.g., section 3).

The transfer equation of the neutrino occupation probability 
$f_{\nu}(\epsilon_{\nu},p,s)$ for a given neutrino energy 
 $\epsilon_{\nu}$ along each ray is given by,
\begin{equation}
\frac{d f_{\nu}(\epsilon_{\nu},p,s)}{d s} = j(\epsilon_{\nu},p,s)(1 - f_{\nu}(\epsilon_{\nu},p,s)) - \frac{f_{\nu}(\epsilon_{\nu},p,s)}{\lambda},
\label{transfer}
\end{equation}
 where $j$ and $\lambda$ is the 
 emissivity and absorptivity via neutrino absorptions and emission by free nucleons 
($\nu_{\text{e}} + \text{n} \rightleftarrows \text{e}^{-} + \text{p}$) 
 (\citet{bruenn_85,ohnishi_1}), which 
 are dominant processes outside the PNSs. The optical depth for those reactions
 are estimated by $\tau_{\nu} = \int_{r}^{\infty} 1/ \lambda$.
 For the sake of simplicity, 
the neutrino scattering and the velocity-dependent terms in the transport 
equation are neglected here.
Although 
 $\bar{\nu}_e$'s are taken into account the hydro simulations,
 we focus only on electron-type neutrinos here for simplicity, since 
 they dominantly contribute to the resulting GWs outside the 
PNSs as will be discussed in section 4 (see also \citet{kotake_gw_sasi}).
We use 16 neutrino energy bins which is logarithmically uniform and covers 0.9 - 110 MeV.  
Along each ray, $f_{\nu}$ is transferred by the line integral.
 When the line integral starts from the surface on the PNS (the lines coming 
 from the PNS in Figure \ref{ray_trace}), 
 we set the initial value of
\begin{equation}
 f(\epsilon_{\nu})
  = \frac{1}{1+\exp(\epsilon_{\nu}/k_{\text{B}}T_{\nu})} \cdot
\frac{1}{4\pi},
\end{equation}
 assuming that the neutrino distribution function at the surface 
is approximated by the Fermi-Dirac distribution with a vanishing chemical potential. 
 Here the neutrino temperature is set to be constant near 
$T_{\nu_{\text{e}}} = 4$ MeV,
 whose values change slightly depending on the input neutrino luminosity.
 Note that these values are constant in time for each model. This is necessary
to realize the steady unperturbed states (e.g., \citet{ohnishi_1}).  
 For the rays that do not hit the PNS, we start the line integral 
from the outer most boundary antipodal to the line of sight, where $f_{\nu}$ is essentially zero.

By a post process, we perform the line integral up to the outer-most boundary for 
each hydro-timestep. 
 The time sampling for the postprocessing is about $\sim$ 1 ms, 
which is sufficient here because the waveforms from neutrinos show 
much slower temporal variation ($\gtrsim 50$ ms) as we will discuss in section 4.
Since the regions outside the PNSs are basically thin to neutrinos (e.g., 
section 4.1), the transport equation is suited to be solved by making 
use of the ray-tracing (long-characteristics) method.
For all the rays, our treatment in equation (11) is identical to neglect the time-retardation, which is equal to 
the light traveling time of neutrinos from the PNS 
to the outer boundary. So the difference of the light traveling time comes 
from the difference in the position of 
 the neutrino emitting regions, which are close to the PNS surface of $\sim$ 50 km. 
Divided by the speed of light, the difference of the time-retardation, 
is less than the order of 0.1 ms, which is negligible for our computation, 
because the hydrodynamical timescale is longer than $\sim$ 10 ms.

 With $f(\epsilon_{\nu},p,s_{\rm out})$, which is obtained by the line integral 
 up to the outer-most boundary, the neutrino energy fluxes 
along a specified direction of ${\bf\Omega}$ can be estimated,
 \begin{equation}
\frac{dl_{\nu}({\bf\Omega},p)}{d\Omega~dS}  = \int f(\epsilon_{\nu},p,s_{\rm out})\cdot(c\epsilon_{\nu}) 
\cdot \frac{\epsilon_{\nu}^2 d\epsilon_{\nu}}{(2 \pi \hbar c)^3}.
\label{flux}
\end{equation}
 By summing up the energy fluxes with the weight of the area 
in the plane perpendicular to the rays, 
we can find $dl_{\nu}/d\Omega$ along a specified direction ${\bf\Omega}$,
\begin{equation}
\frac{dl_{\nu}({\bf\Omega})}{d\Omega} = \int \frac{dl_{\nu}({\bf\Omega},p)}{d\Omega dS}~dS = 
\int_{p_{\rm in}}^{p_{\rm out}} dp~2 \pi p~ \frac{dl_{\nu}({\bf\Omega},p)}{d\Omega dS}.
\label{final}
\end{equation}
Repeating the above procedures, $dl_{\nu}({\bf\Omega})/d\Omega$ 
 can be estimated for all the directions.

\begin{figure}[hbt]
\epsscale{1.1}
\plottwo{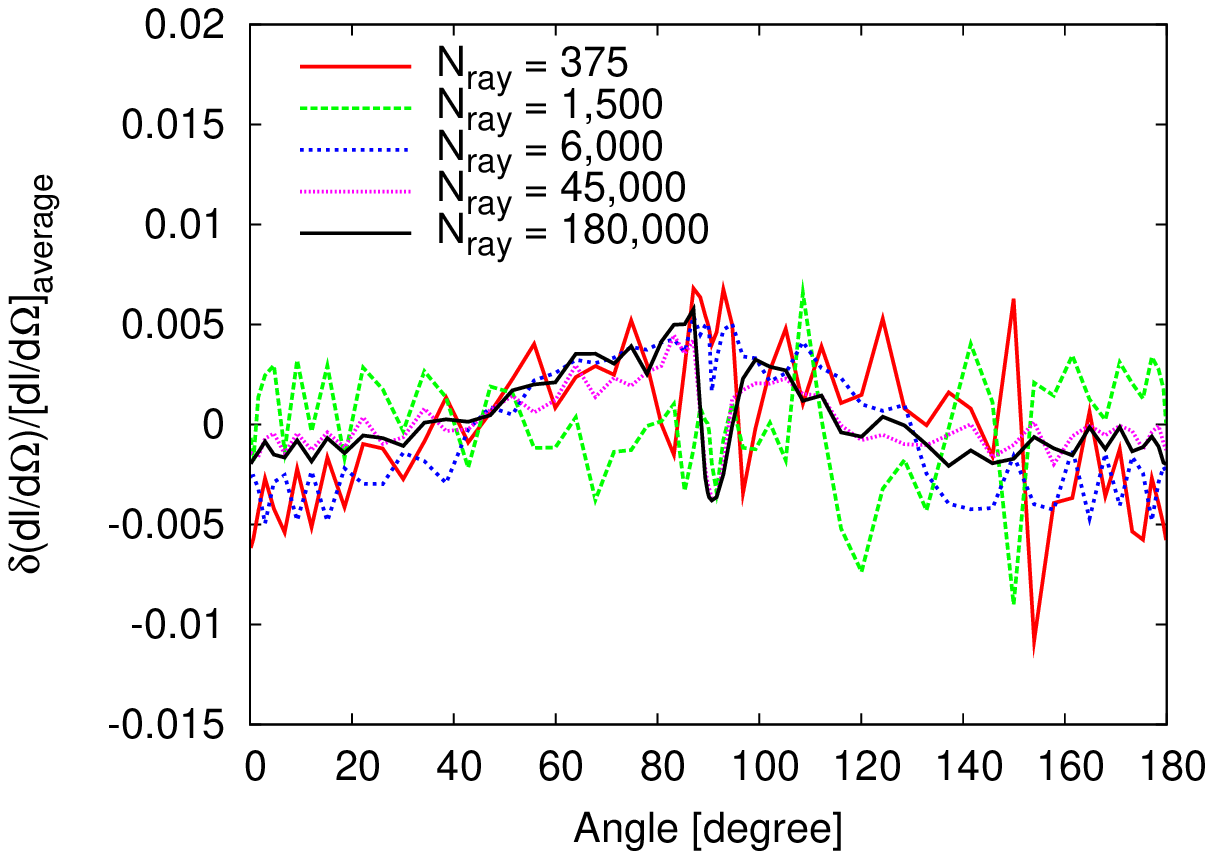}{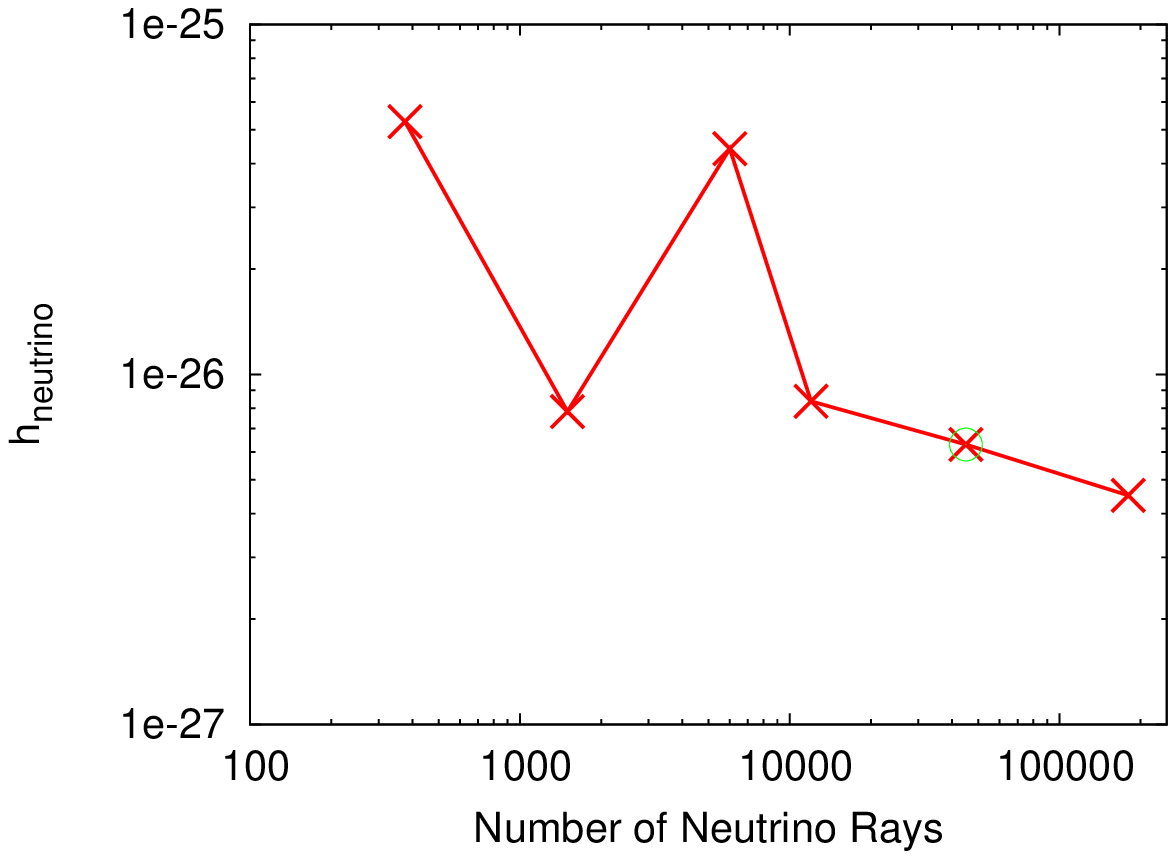}
\caption{Numerical tests, in which the ray-tracing calculation is done for a spherical 
 medium using data at $t$ = 294 ms for model A. Left panel 
 shows the deviation of $dl_{\nu}({\bf\Omega})/d\Omega$ 
from its angle-averaged value as a function of the angle measured from the polar axis
 for different number of the neutrino rays (:$N_{\rm ray}$) (see text for more details). 
Right panel shows the resulting GW amplitudes, which should formally vanish. The green  circle represents for $N_{\rm ray} = 45,000$, which we employ 
for our actual ray-tracing calculation. Note in the right panel that the supernova is assumed to be 
located at a distance of 10 kpc. }
\label{f2_added}
\end{figure}

\begin{figure}[hbt]
\epsscale{0.5}
\plotone{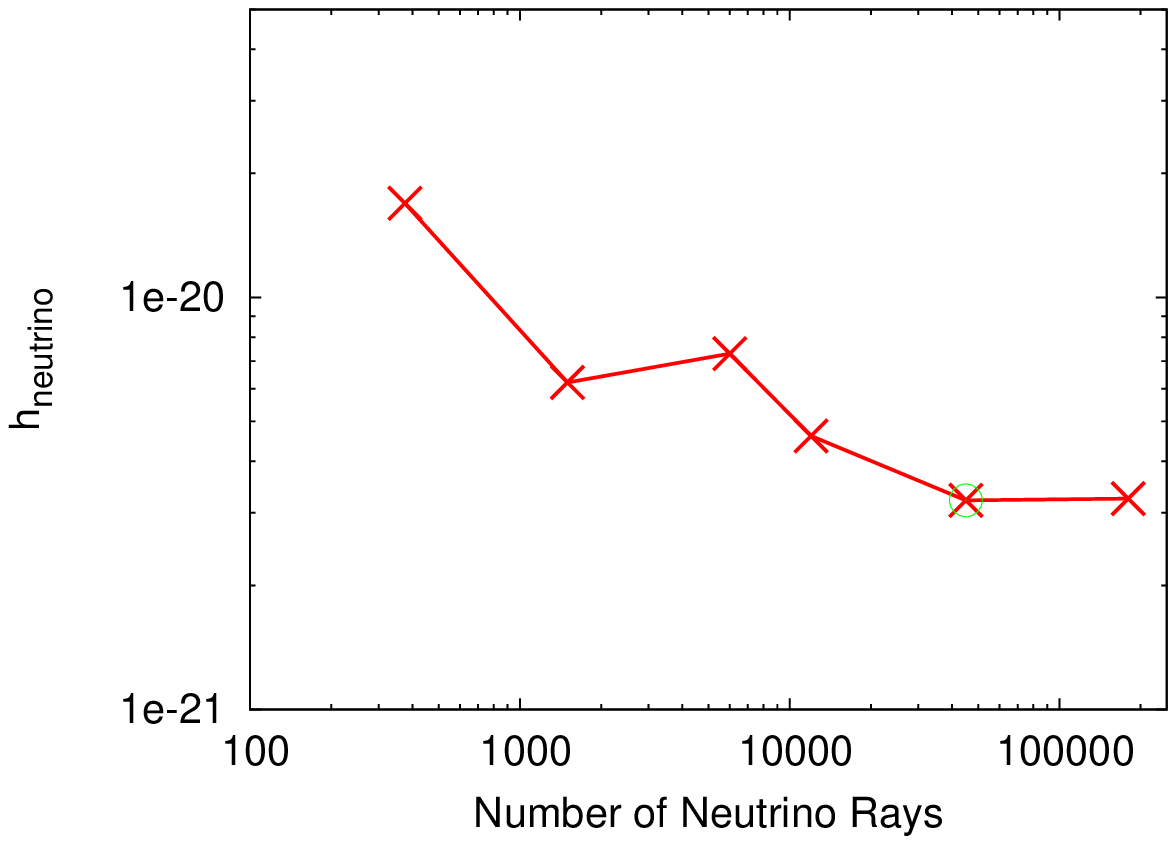}
\caption{Same as the right panel of Figure \ref{f2_added} but for a deformed
 medium. Here we artificially impose a global quadrupole deformation ($\ell = 2$) to 
the matter distribution. It shows the numerical convergence of the 
GW amplitudes for $N_{\rm ray} = 45,000$.
 }
\label{f4_added}
\end{figure}

To verify the newly developed numerical code, we calculate 
 $dl_{\nu}({\bf\Omega})/d\Omega$ in a spherical medium as a test calculation.
The left panel of Figure \ref{f2_added} shows how the deviation 
of $dl_{\nu}({\bf\Omega})/d\Omega$ 
from its (polar)angle-averaged value, indicated by 
$\delta(dl_{\nu}/d\Omega) \equiv dl_{\nu}/d\Omega - 
 (dl_{\nu}/d\Omega)_{\rm average}$ in the figure, 
changes with the number of the neutrino rays. With increasing the number of 
 the rays, the deviation is shown to be smaller especially near the polar 
 regions (near $0^{\circ},~180^{\circ}$), 
while the deviation near the equator (near $90^{\circ}$) 
is nearly converged, but with relatively larger errors than for the other direction.
 This is because the angular grid size near the equatorial belt becomes
  larger for the spherical coordinate system taken in our simulation. Right panel 
of Figure \ref{f2_added} shows the resulting GW amplitudes for the 
different number of the 
rays. It is noted that the amplitudes vanish formally 
for the isotropic configuration. 
 From the restriction of the computational time, 
we choose to set 45,000 rays for our actual ray-tracing calculation (green circle 
in the right panel). With this  
 choice, the numerical errors are suppressed to be an order of $10^{-27}$, which is
 typically 4-5 orders of magnitudes smaller than typical GW amplitudes 
obtained in the models computed here. We have also checked the numerical 
convergence for a deformed matter configuration (see Figure \ref{f4_added}).

\section{Construction of 2D Exploding Models \label{sec3}}
As in \citet{kotake_gw_sasi}, we assume that the explosion is powered by 
the neutrino-energy deposition
 between the PNS and the standing accretion shock, probably one of the most promising way to 
blow up massive stars (e.g., \citet{marek}).
To trigger the explosions, we employ the so-called light-bulb approximation 
 (see \citet{jankamueller96,ohnishi_1} for details) 
and adjust the neutrino luminosities from the
PNSs, which was found to work well in 2D, capturing the main features obtained 
by more realistic simulations (e.g., \citet{scheck_04,scheck}).
 It should be noted again that we focus on the GWs generated by the anisotropies 
 associated with the growth of SASI, which develops outside the PNS.
 Therefore, the oscillations of the PNSs \citep{burr_new} 
 and the resulting efficient gravitational emission \citep{ott_new} or 
 the enhanced neutrino emission near the equator of the PNS observed 
 in \citet{marek_gw},
  cannot be treated in principle here. 
 


Our hydrodynamic code is based on the ZEUS \citep{stone} and 
 major modifications for the supernova simulations 
are already described in \citet{kotake,ohnishi_1}. 
The computational grid is comprised of 300 logarithmically spaced
radial zones to cover from the absorbing inner boundary of 
$\sim 50 ~{\rm km}$ to the outer boundary of $2000 ~{\rm
km}$, and 60 polar ($\theta$) uniform mesh points
(see for the resolution tests in \citet{kotake_gw_sasi}).

The initial condition is a spherically symmetric steady accretion
flow through a stalled shock wave to a PNS \citep{yamasaki}, which is also
 provided in the same manner of \citet{ohnishi_1}.
In constructing the initial conditions, we assume
 a fixed density $\rho_{\rm in} = 10^{11}$~g~cm$^{-3}$ at the inner boundary.
And the initial mass accretion rates and the initial 
mass of the central object are set 
to be $\dot{M} = 1~M_{\odot}$~s$^{-1}$ and $M_{\rm in} = 1.4~M_{\odot}$,
respectively. As in \citet{kotake_gw_sasi}, the self-gravity of matter 
in the accretion flow is ignored. To induce
non-spherical instability, we add random velocity perturbations of 
less than 1 $\%$ of  the unperturbed velocity.
 At the outer boundary, we adopt the fixed boundary condition consistent 
with the initial condition. On the other hand, the absorbing boundary 
is used at the inner boundary. 
We assume that the neutrino flux from the PNSs can be approximated by 
black-body emission. 
 
 Pushed by the evidences that 
the SASI-induced explosions are favorable for explaining the observed 
quantities such as the synthesized elements of SN1987A \citep{kifo}, 
the pulsar kicks \citep{scheck_04} and spins \citep{blondin07a,iwakami08_2}, 
  we mainly focus on the gravitational radiation in the models tuned 
to produce explosions.
By changing the electron neutrino luminosity at the surface of the PNS
in the range of $L_{\nu_e} = 6.4 - 6.8 \times 10^{52}(10~{\rm foe})$~erg~s$^{-1}$, 
 we construct three exploding and one non-exploding models (see Table \ref{table1}).
Except for model D, we can observe the continuous increase of the 
average shock radius with the growth of SASI, reaching the outer 
boundary of the computational domain with the
explosion energy of $\sim 10^{51}$ erg. Until this moment, we run the 
simulations (e.g.,  $\Delta t$ in the table), while in model D, 
we terminated the simulation at about 800 ms, not seeing the increase of 
the shock radius.

\begin{deluxetable}{cccccc}
\tabletypesize{\scriptsize}
\tablecaption{Model Summary \label{table1}}
\tablewidth{0pt}
\tablehead{
\colhead{Model}
 & $L_{{\nu}_e}$ ($10^{52}$ erg/s)
 & \colhead{$\Delta t~~({\rm ms})$}
 & \colhead{$h_{\nu, \rm fin}$ ($10^{-22}$)}
 & \colhead{$|h_{\rm tot, max}|$ ($10^{-22}$)}
 & \colhead{$E_{{\rm GW},\nu}$ ($10^{-12} M_{\odot} c^2$)}}
\startdata
A& 6.8 & 509      & 8.7  & 7.7 &  0.44 \\
B& 6.7 & 570      & 2.2  & 9.1 &  1.32\\ 
C& 6.6 & 740      & 6.1  & 8.0 &  1.39\\
D& 6.4 & 800      & 4.8  & 6.1 &  0.49\\
\enddata
\tablecomments{%
$L_{{\nu}_e}$ denotes the input luminosity.
$\Delta t$ represents the simulation time. 
$h_{\nu, \rm fin}$ and $h_{\rm tot, max}$ 
represents the amplitudes of the neutrino-originated GWs at the end 
of the simulations and the maximum amplitudes (neutrino + matter) during the simulation 
time.
$E_{{\rm GW,}\nu}$ is the radiated energy in the form of the 
 neutrino GWs in unit of $M_{\odot} c^2$. 
Note that the supernova is assumed to be located at a distance of 10 kpc. }
\end{deluxetable}

\begin{figure}[hbt]
\epsscale{1.1}
\plottwo{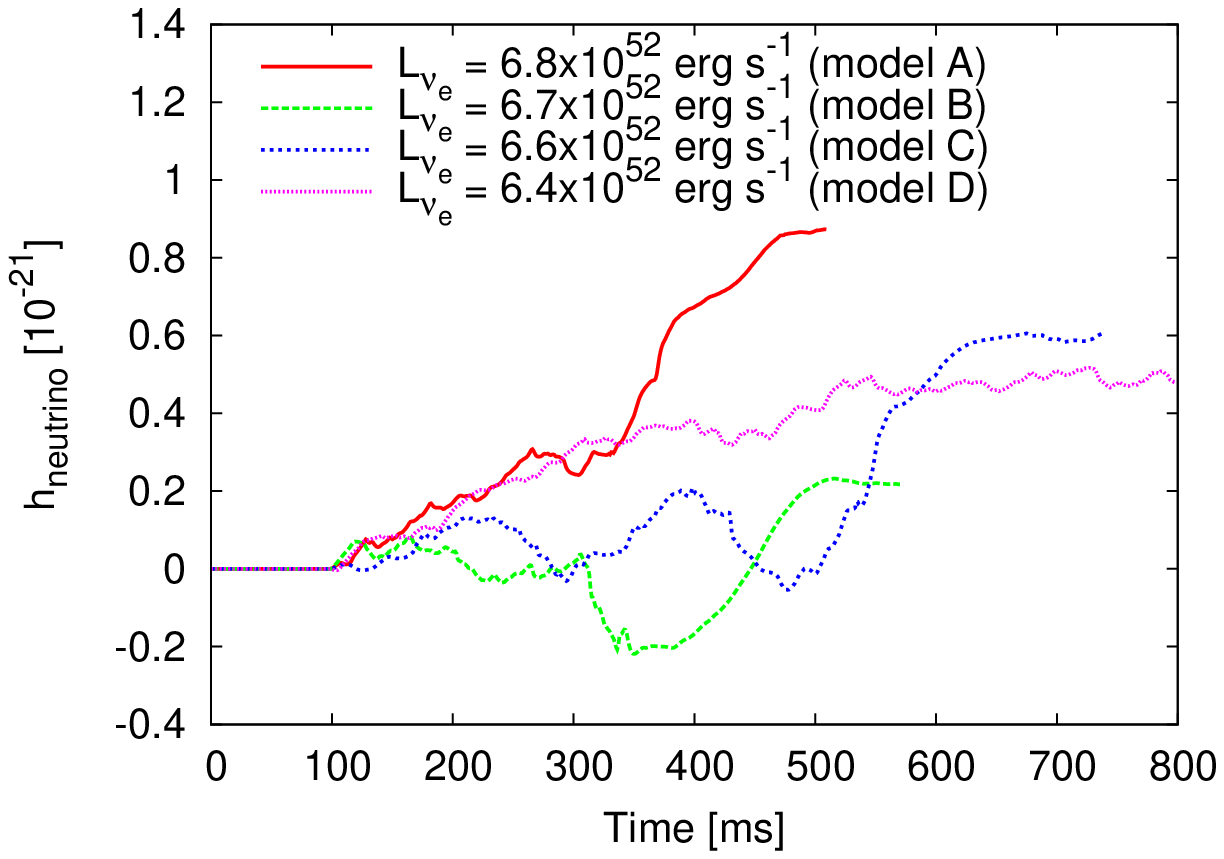}{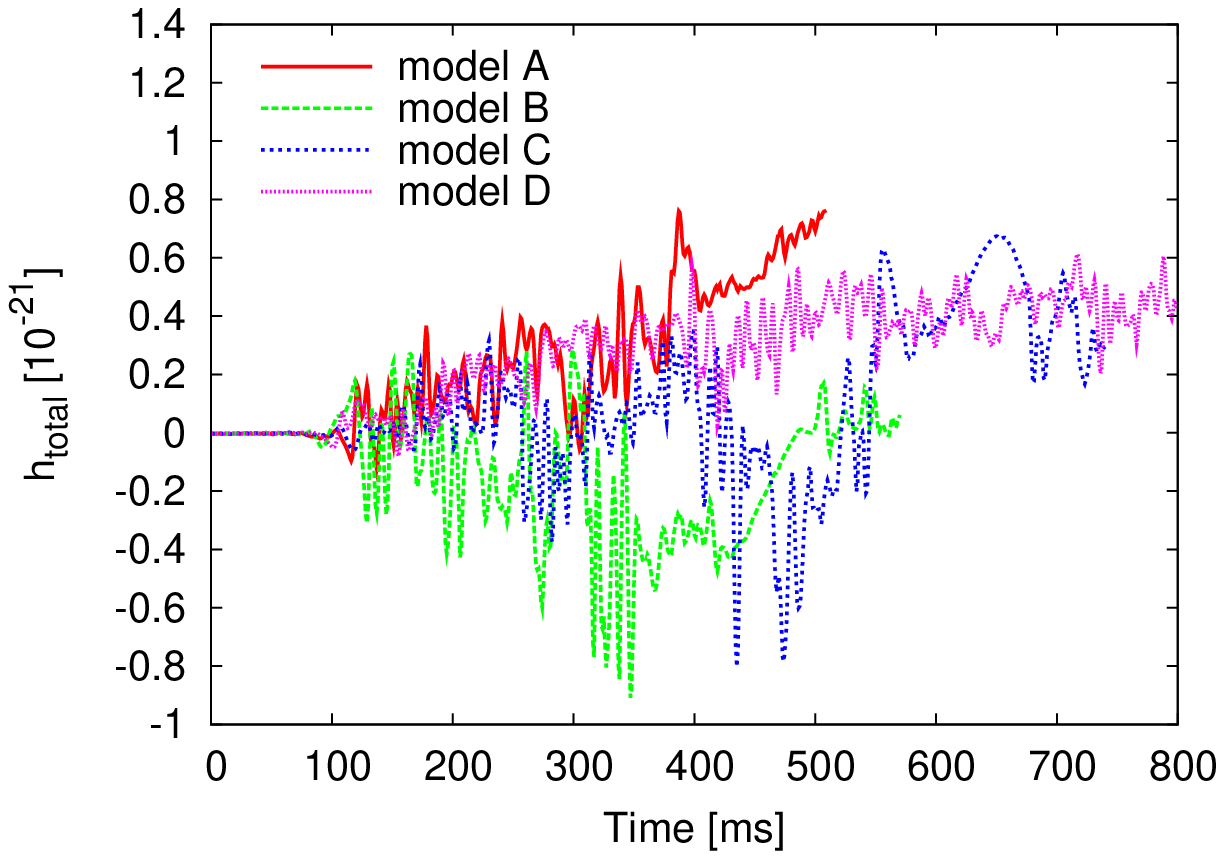}
\caption{Gravitational waveforms only from anisotropic neutrino emission (left)
 and from the sum of neutrinos and matter motions (right).
 The time is measured from 
the epoch when the neutrino luminosity is injected from the surface of the 
neutrino sphere. In all the computed models, SASI enters to the non-linear regime 
at about $100$ ms, simultaneously making the amplitudes deviate from zero. 
Note that the supernova is assumed to be located at the distance of 10 kpc.}
\label{f3}
\end{figure}

\begin{figure}[hbt]
\epsscale{1.1}
\plottwo{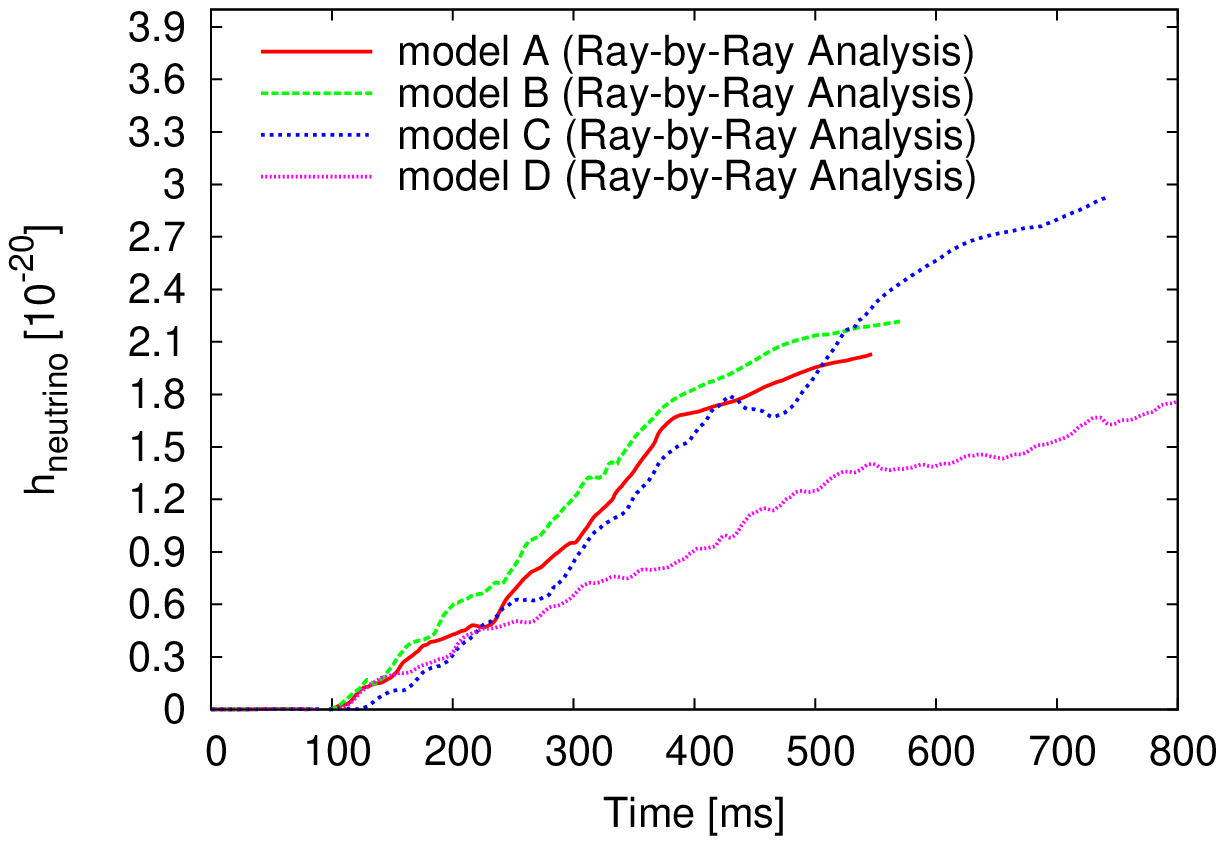}{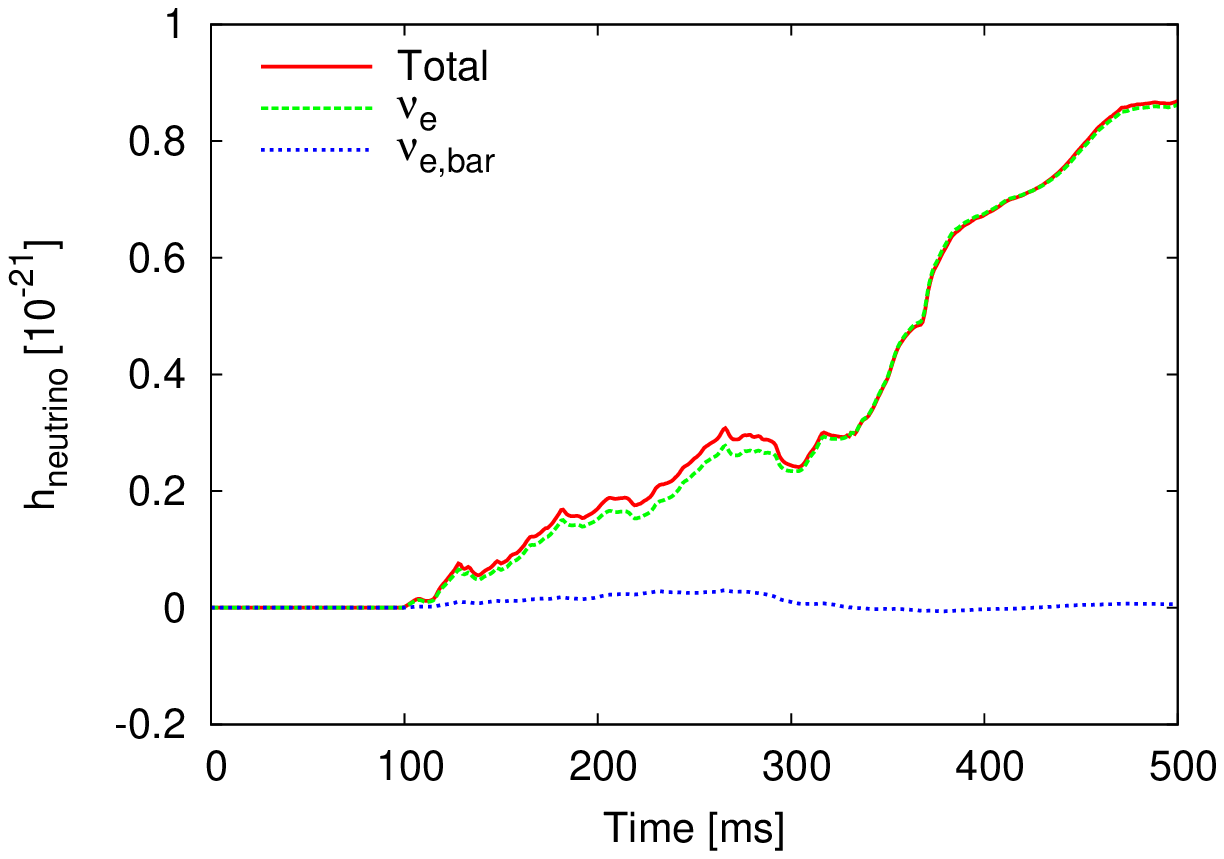}
\caption{Gravitational waveforms from anisotropic neutrino emission estimated
 by the ray-by-ray analysis of \citet{kotake_gw_sasi}, showing a monotonic
 increase of the amplitudes with time (left). Right panel is the same as the 
 left panel of Figure \ref{f3} for model A but with the contribution of anti-electron type neutrinos ($\bar{\nu}_{e}$) to 
the waveform (indicated by ``Total''), showing that the dominant contribution to the GW amplitudes
 comes from electron-type neutrinos ($\nu_e$). The supernova is assumed to be located 
at the distance of 10 kpc.}
\label{f3_added}
\end{figure}
\section{Results\label{sec4}}

 The left panel of Figure \ref{f3} shows the GW amplitudes
contributed from anisotropic neutrino emission for 
different luminosity models. Comparing the right panel, which shows the 
  total amplitudes (neutrino + matter),
 it can be seen that the gross structures of the waveforms are 
 predominantly determined by the neutrino-originated GWs with the slower temporal 
 variations ($\gtrsim 50$ ms), to which the GWs from matter motions 
 with rapid temporal variations ($\lesssim 10$ ms) are superimposed. 

On the other hand, the waveforms shown in the left panel 
 of Figure \ref{f3_added} 
 are estimated simply by summing up the local 
neutrino cooling rates outside the PNSs with the ray-by-ray assumptions
 \citep{kotake_gw_sasi} (see also Figure 2 in the paper).
They are so much different from the ones obtained here.
In contrast to the monotonic increase of the amplitudes with time, 
 the waveforms here exhibit more variety, showing large negative growth at some 
epochs. And the GW amplitudes from neutrinos become more than 
one-order-of magnitude smaller than the previous estimation (compare Figures
 \ref{f3} and the left panel of \ref{f3_added} noting the different vertical scales).
 On the other hand, similarities between them are that the waveforms 
from neutrinos have generally a positively growing feature with time, 
 and also that the electron-type neutrinos dominantly contribute to the wave amplitudes than for the 
anti-electron type neutrinos (right panel of Figure \ref{f3_added}, see also 
 Figure 3 in \citet{kotake_gw_sasi}).
In the following, we first analyze this positive trend based on the 
 ray-tracing calculations. And in section \ref{naga},  
we clarify the reason for the negative growth, contributing to make the amplitudes 
much smaller than previously estimated.  

\begin{figure}[hbt]
\epsscale{1.0}
\plottwo{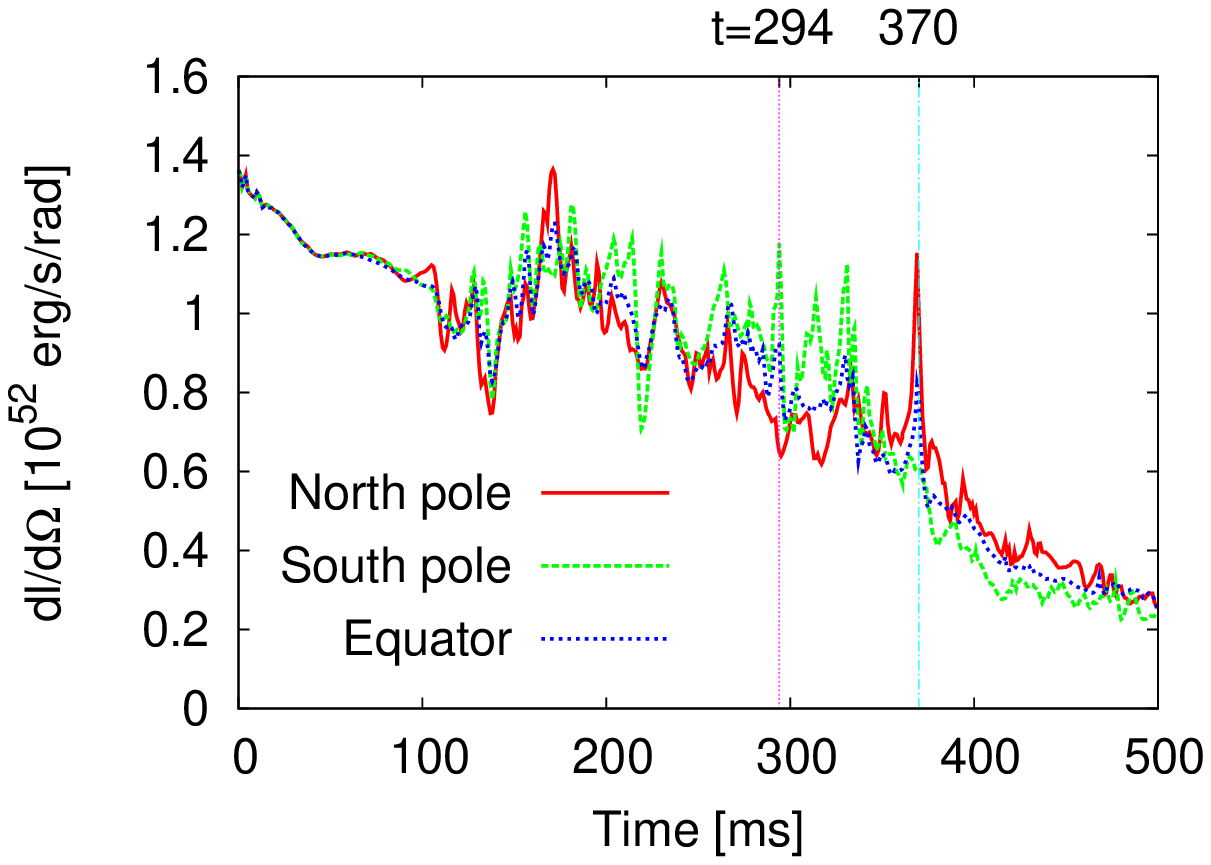}{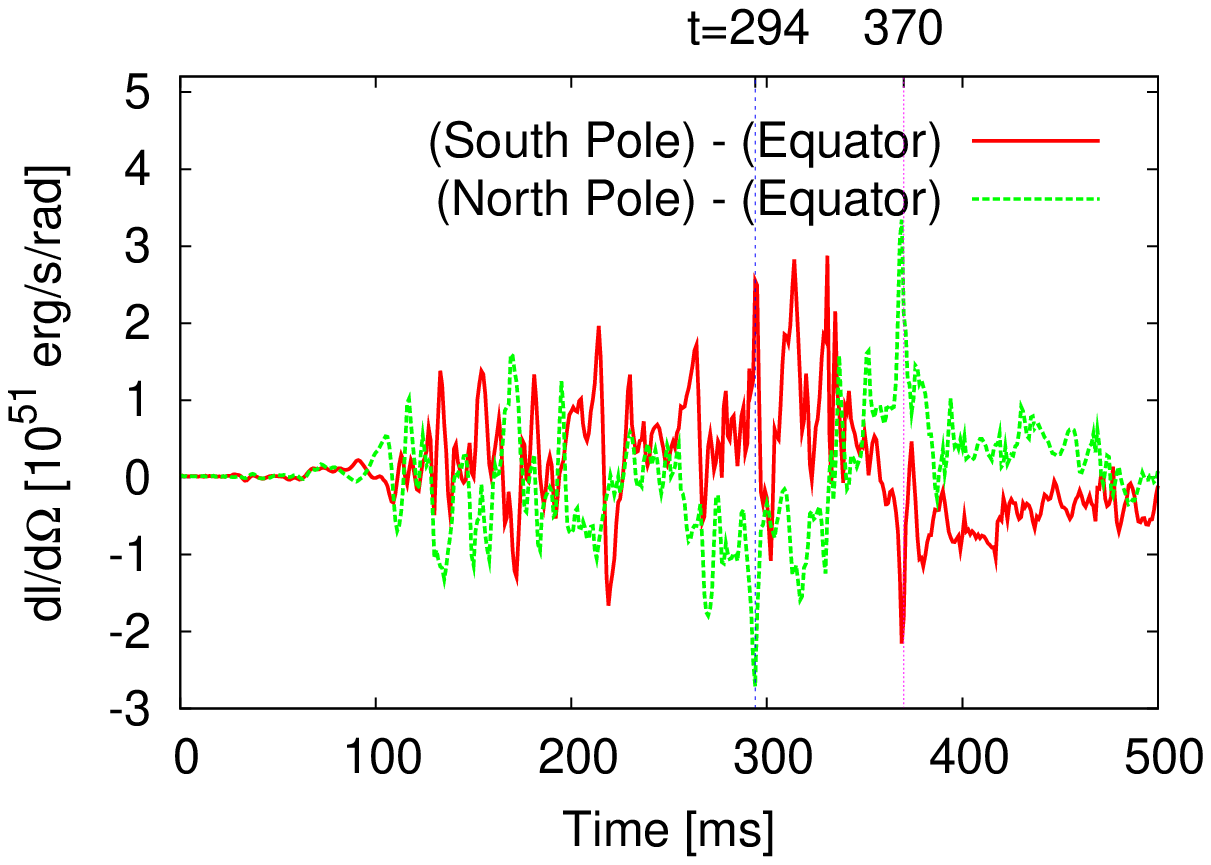}
\caption{The directional dependent neutrino luminosity: $dl_{\nu}/d\Omega$ for 
model A, in the vicinity of the north pole, the equator, and the south pole 
(left panel), and their differences from the equator 
(right panel). Vertical lines represent the epochs of $t$ = 294 and 370 ms, respectively 
(see text for details).}
\label{f4}
\end{figure}

\begin{figure}[hbt]
\epsscale{0.8}
\plotone{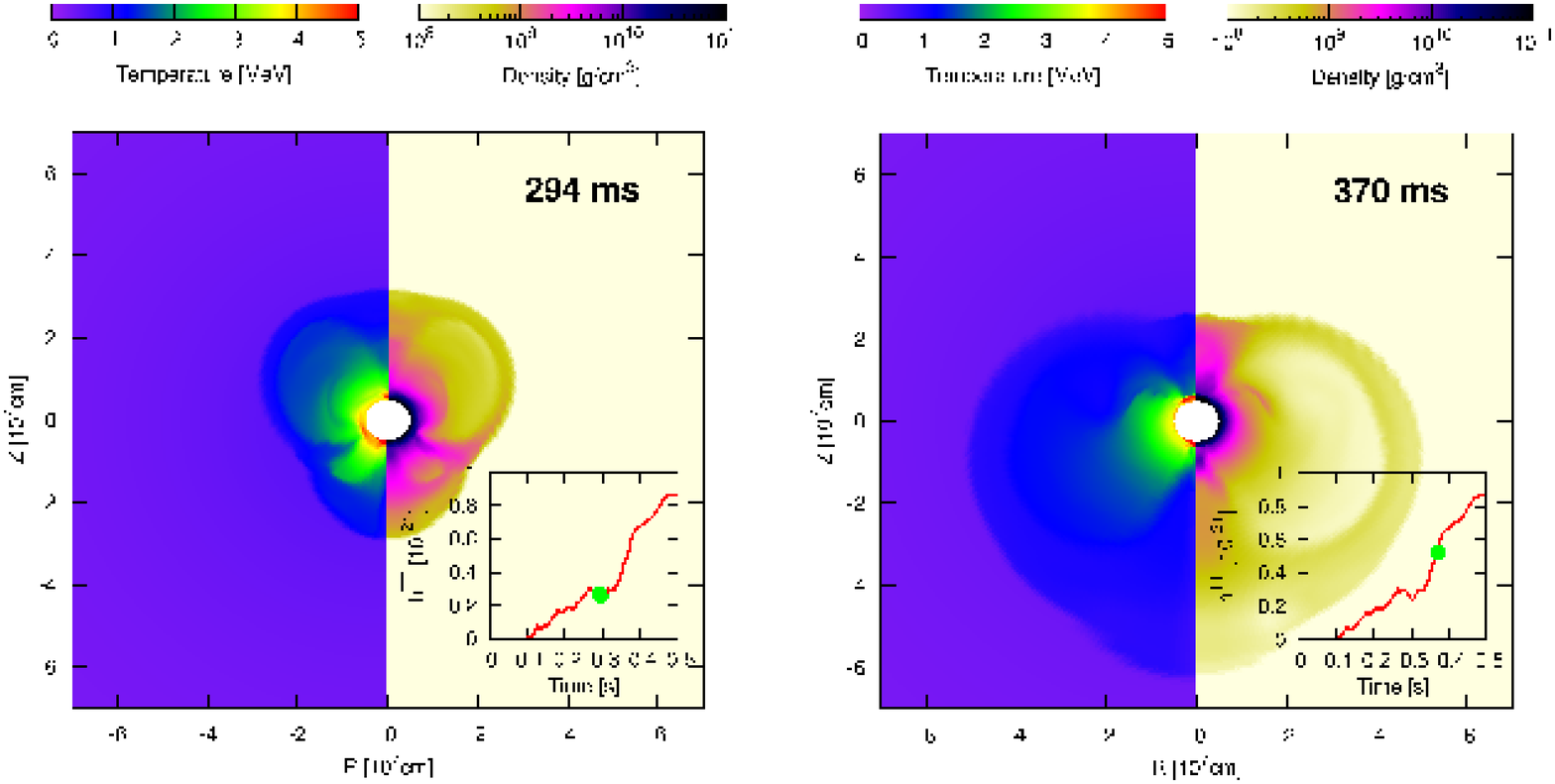}
\caption{Temperature- (the left-half of each panel) and density- 
(the right-half) distributions in the meridian section for model A. The insert of 
each panel shows the neutrino GW amplitudes, in which the green point indicates 
the time of the snapshot. The central region colored by white (50 km in radius) 
 represents the inner boundary of our computation.}
\label{f5}
\end{figure}

\subsection{Positively growing features  \label{posi}}
To see the positively growing features clearly, we choose model A as a reference.
The left panel of Figure \ref{f4} shows $dl_{\nu}/d\Omega$ in the vicinity of the 
 north pole ($\theta=0$), the equator ($\theta=\pi/2$), and the south pole
($\theta=\pi$). In the following, we focus on the two epochs of $t$ = 294 and 370 ms, 
when the neutrino emission in the south pole dominate over the ones 
in the north pole and vice versa.
The right panel of Figure \ref{f4} shows that the dominance of the 
neutrino emission in the north and south poles are closely anti-correlated. 
This is the consequence of the low-mode nature of SASI, here of $\ell=1$.
 In fact, the right panel of Figure \ref{f5} shows that at 294 ms, 
the blob encompassing the regions inside the stalled shock is moving from the southern 
to the northern hemisphere, 
leading to the compression of the matter in the south hemisphere, 
which is vice versa at 370 ms (right panel).

\begin{figure}[hbt]
\epsscale{0.9}
\plotone{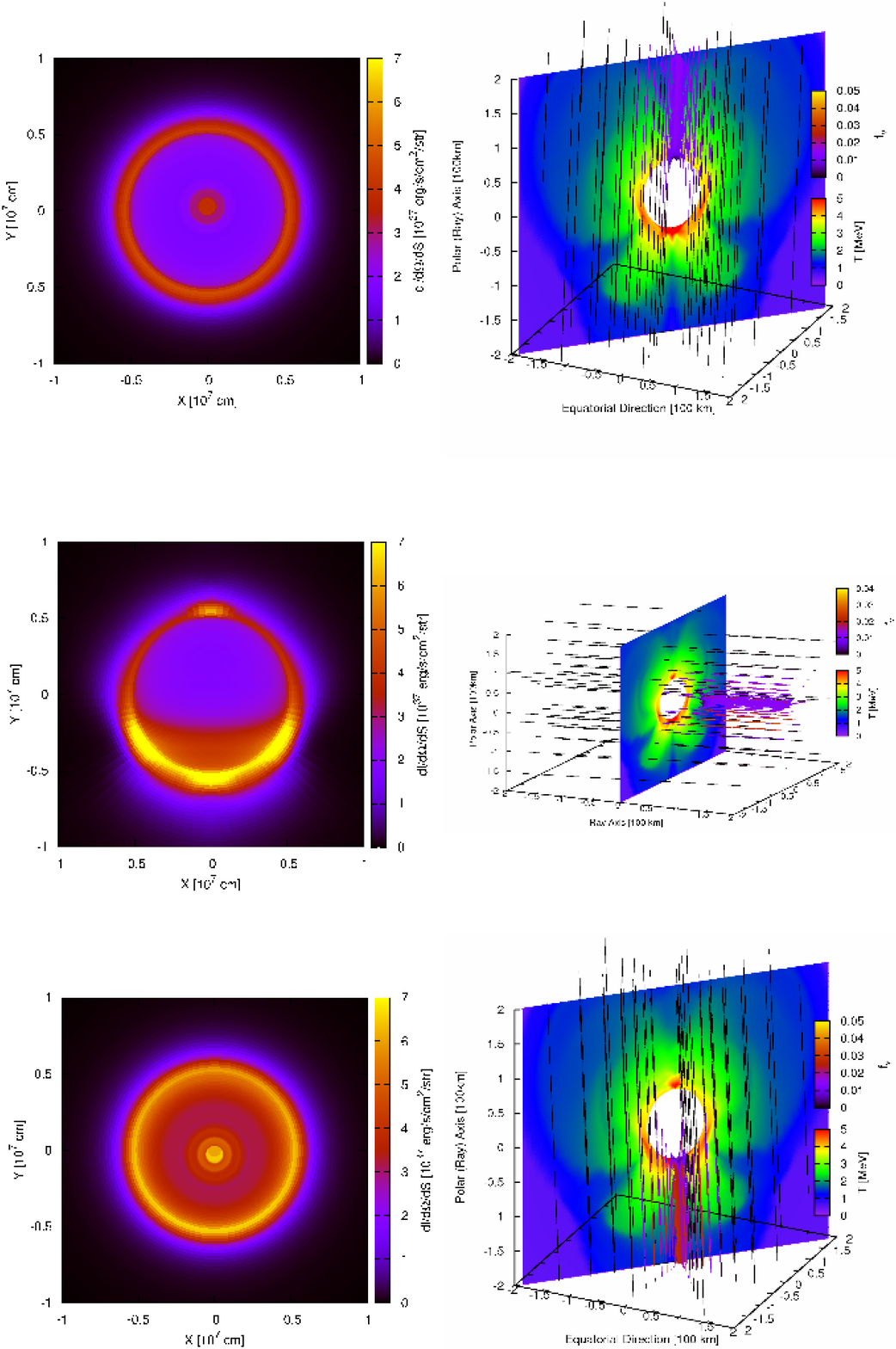}
\caption{Various properties obtained by the ray-tracing calculation 
at $t=294$ ms for model A. Right panels show the rays 
(colored by the value of $f_{\nu}$ for $\epsilon_{\nu} = 14$ MeV), which are 
superimposed on the temperature distributions 
in the meridian section. Left panels show the neutrino energy fluxes of
$dl_{\nu}/({d\Omega dS})$ (equation(\ref{flux})) seen from the 
 northern hemisphere (top), the equator (middle), and the southern hemisphere (bottom), respectively. 
 }
\label{f6}
\end{figure}

Figure \ref{f6} shows various properties obtained by the 
ray-tracing calculation for $t=294$ ms.
Top, middle, and bottom panel, is the case seen from the northern hemisphere, 
the equator, 
and the south hemisphere, respectively.
Comparing the right panels, we can see that the rays with highest $f_{\nu}$, come 
from the high-temperature regions near the vicinity of 
the south pole (bottom right panel). 
As mentioned above, this is due to the compression of matter near the south pole, 
produced by the sloshing SASI, moving from the south to the northern hemisphere.

\begin{figure}[hbt]
\epsscale{1.0}
\plotone{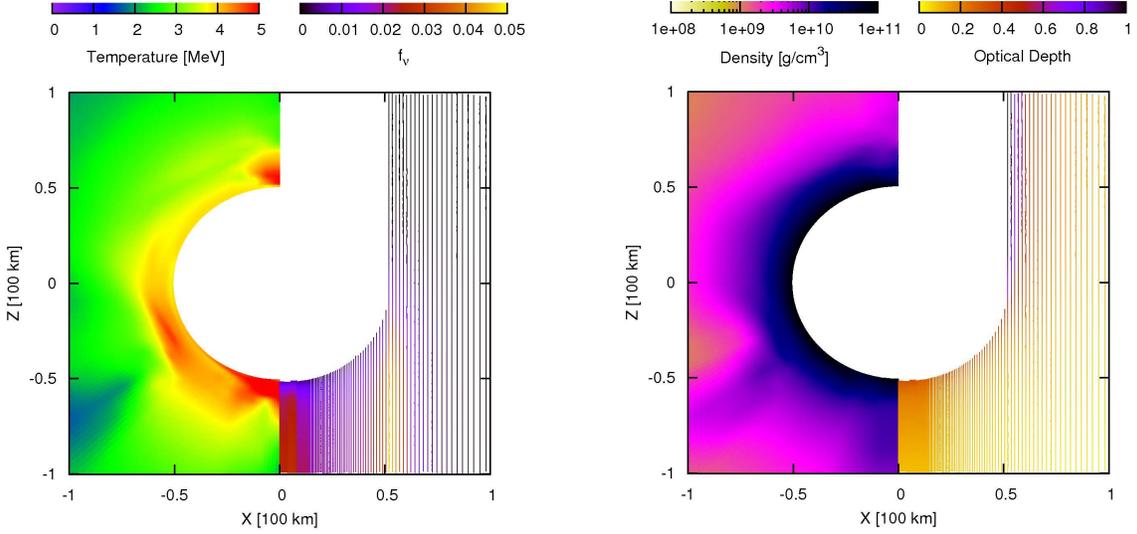}
\caption{Left panel is an enlargement of the bottom right panel of Figure \ref{f6} 
near the central PNS (white circle). Right panel depicts 
the optical depth of neutrinos at $\epsilon_{\nu} = 14$ MeV (right-hand side) with the
 density distributions (left-hand side). The opacity sources are taken to be 
 the neutrino emission/absorption as mentioned in section \ref{sec2.2}. 
It can be seen that high values of $f_{\nu}$ come from the high temperature regions in the vicinity of the PNS, 
where are thin to neutrinos with the optical depth being smaller than $2/3$.}
\label{f7}
\end{figure}

In the left panels of Figure \ref{f6}, it is noted that due to the axisymmetry, 
the distributions of the neutrino energy fluxes seen from the north 
(top left) and south hemispheres (bottom left) have the circumferential structures. 
 The images of the energy fluxes may seem like an annular 
eclipse, which we explain as follows. Here for clarity, let's
 consider two rays of A and B. 
The ray A is coming a point from ($X=0, Y=0$) in the middle left panel, 
which in the middle right panel, is equal to the point along the line 
perpendicularly threading the center of the colored plane, but is positioned
 at 50 km away from the plane to the observer (:right direction in the panel). 
The ray B is coming from a point from 
($X=0, Y\approx - 50 $km) in the middle left panel, which is equal to the point, 
 near the southern point on the colored plane in the middle right. The ray 
 B travels to the observer, experiencing longer the high-temperature regions
 near the surface of the PNS. This makes 
 the eclipse-like shining near the edge of the PNS. 
 The left panel of Figure \ref{f7} is an enlargement of the bottom right panel 
of Figure \ref{f6} near the PNS, which clearly shows the rim-shining near 
$X = 50$ km (compare the right-hand 
side with the bottom left panel of Figure \ref{f6}). The right-hand side in the 
right panel of Figure \ref{f7} depicts the optical depth of neutrinos at 
$\epsilon_{\nu} = 14$ MeV. It is seen that the high values of $f_{\nu}$ come from 
the higher temperature regions 
 in the vicinity of the southern part of the PNS (left panel), where are thin to the 
neutrinos with the optical depth being smaller than $2/3$ (right panel). 
Such an optically thinness is favorable for the ray-tracing calculation here 
as discussed in section \ref{sec2.2}.
 
\begin{figure}[hbt]
\epsscale{1.0}
\plottwo{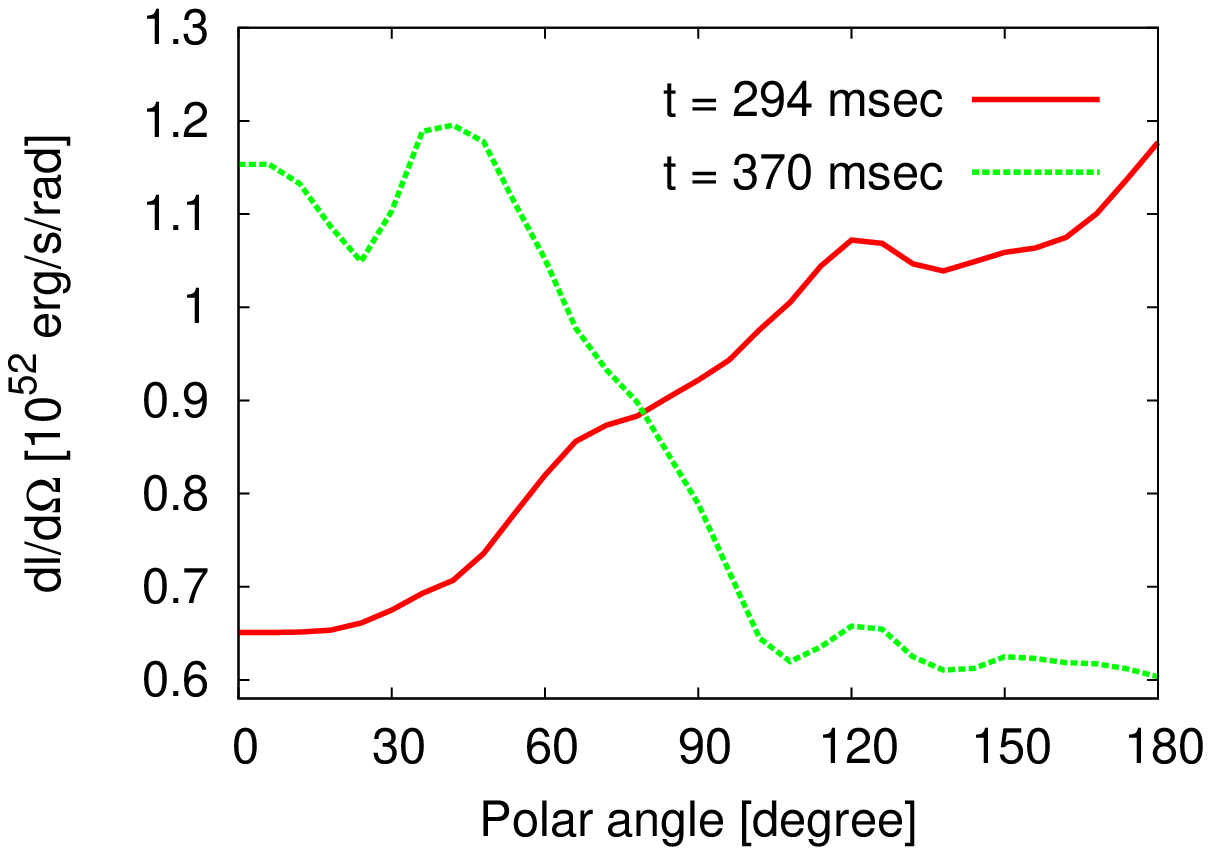}{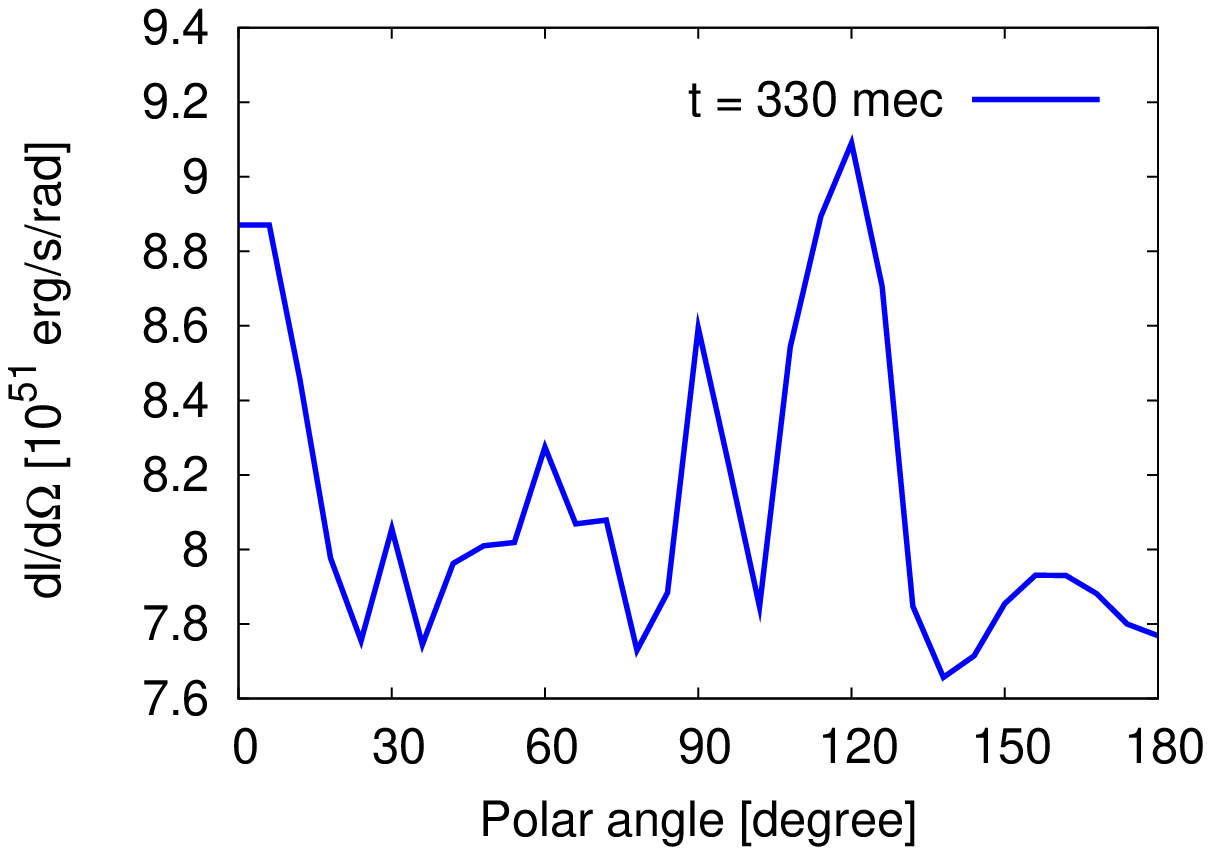}
\caption{The directional dependent neutrino luminosities of $dl_{\nu}/d\Omega$ as 
a function of the polar angle for model A (left panel) at $t = 294$ and $t = 370$ ms, 
 and model B (right panel) at $t = 330$ ms.}
\label{f8}
\end{figure}

\begin{figure}
\epsscale{1.0}
\plotone{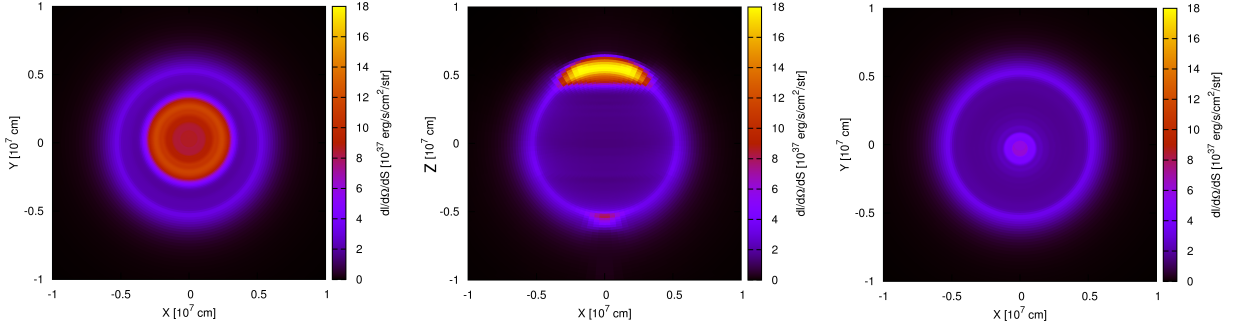}
\caption{Neutrino energy fluxes of $dl_{\nu}/({d\Omega dS})$ (equation(\ref{flux})) of 
 model A at $t=370$ ms, seen from the northern hemisphere (left), the equator (middle), 
and the southern hemisphere (right), respectively.}
\label{f9}
\end{figure}

 Among the left panels of Figure \ref{f6}, 
 the highest values of the local neutrino energy fluxes (:$dl_{\nu}/({d\Omega dS})$) 
come from the brightly shining southern poles seen from the equator (middle panel). 
However by summing up them with the weight of the area (e.g., $dS$ in equation (\ref{final})), 
$dl_{\nu}/d\Omega$ are shown to become largest seen from the southern hemisphere 
  (the red line in the left panel of Figure \ref{f8}). 
The similar argument is true for 
 $t = 370$ ms (Figure \ref{f9}), while the neutrino luminosities seen from the northern
 hemisphere are higher than the ones from the southern hemisphere (the
 green line in the left panel of Figure \ref{f8}).
 Noting again that $\Phi(\theta^{'})$ in equation (\ref{tt}) are positive near 
the north and south polar caps, the dominance of the anisotropic neutrino luminosities 
in the vicinity near the north and south poles, makes the 
positively growing feature in the resulting GWs.

\begin{figure}[hbt]
\epsscale{0.5}
\plotone{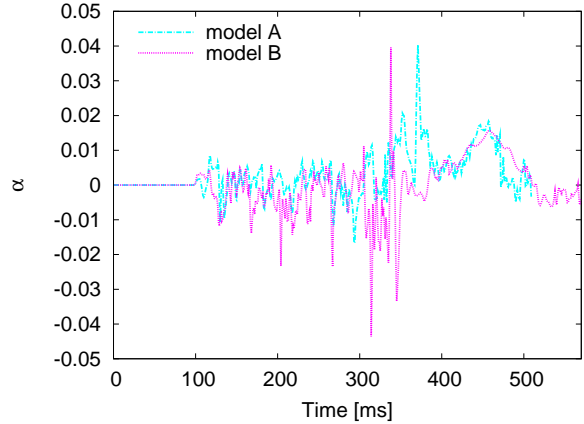}
\caption{Time evolution of the neutrino anisotropy parameter (:$\alpha$ in equation
 (\ref{alpha})) for models A and B. 
$\alpha$ keeps positive value with time in the later phase ($\gtrsim 400$ ms) 
 when the low-modes explosion is triggered by SASI 
along the symmetry axis.}
\label{f7_added}
\end{figure}

Figure \ref{f7_added} shows time evolution of a neutrino anisotropy parameter 
defined in \citet{muyan97},
\begin{equation}
\alpha(t) =
\frac{1}{L_{\nu}(t)}\int_{4\pi}~d\Omega'
~\Phi(\theta')~\frac{dl_{\nu}(\Omega',t')}{d\Omega'},
\label{alpha}
\end{equation}
 where $L_{\nu}(t)$ is the total neutrino luminosity $L_{\nu}(t) =\int_{4\pi} d\Omega' {dl_{\nu}(\Omega',t')}/{d\Omega'}$. This quantity is useful to see how the temporal 
changes of the neutrino anisotropy have impacts on the GW amplitudes. 
Either for models A or B, it can be seen that 
$\alpha$ keeps positive value with time in the later phase ($\gtrsim 400$ ms).
 This coincides with the epoch when the low-modes explosion is triggered by SASI along 
the symmetry axis, which is also helpful for understanding 
the reason of the positive growth (compare the left panel of Figure \ref{f3}). 

\begin{figure}
\epsscale{.8}
\plotone{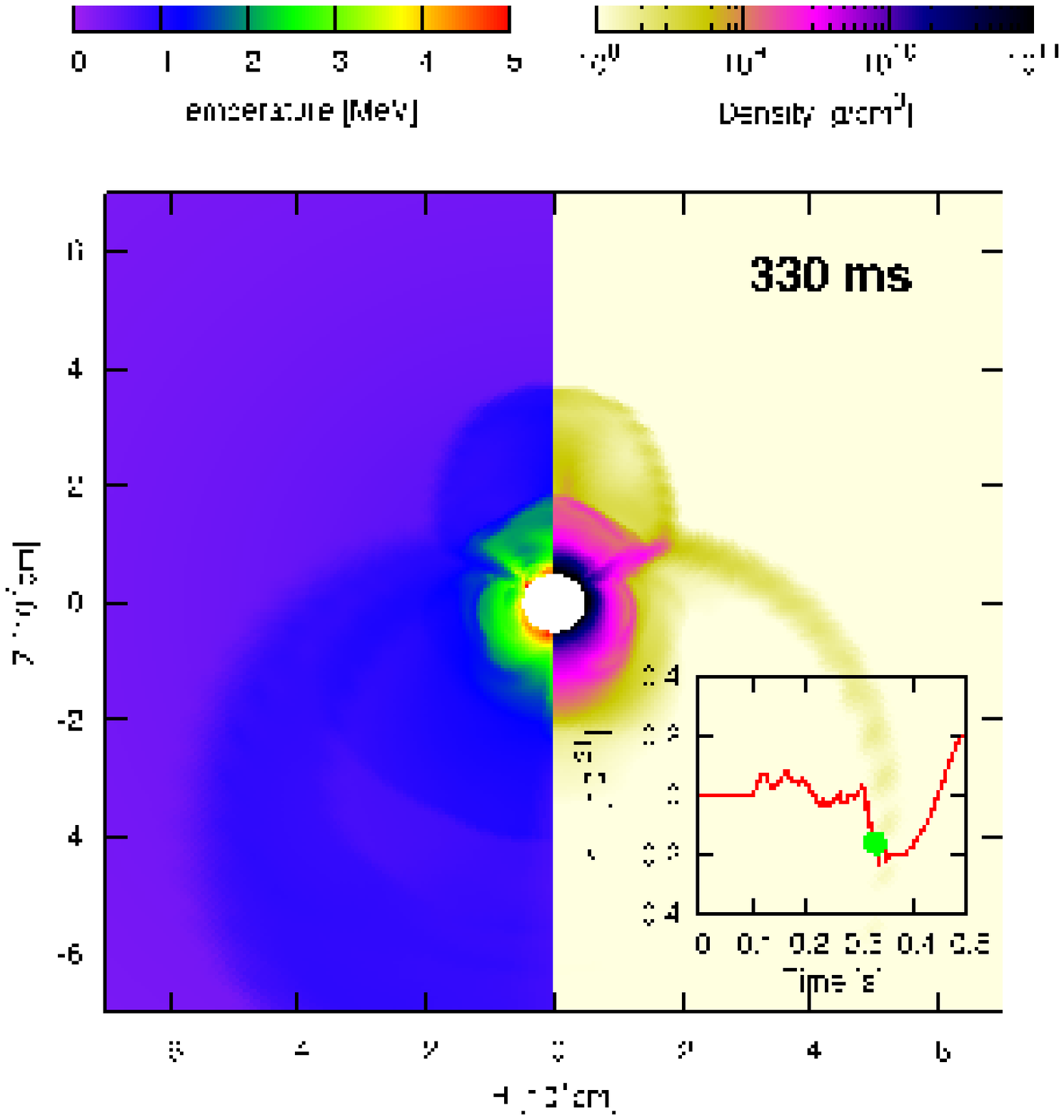}
\caption{Same as Figure \ref{f5} but for model B at $t = 330$ ms.}
\label{f10}
\end{figure}

\subsection{Negatively growing features \label{naga}}

 Now we move on to discuss large negative growth of the GW amplitudes seen in the waveforms
 (e.g., models B and C in Figure \ref{f3}), which 
 is newly found by the ray-tracing calculation 
(compare Figure 2 of \citet{kotake_gw_sasi}). 
Here we choose a snapshot at $t=330$ ms of model B as a reference, 
which shows the steep negative growth (see the insert of Figure \ref{f10}).

\begin{figure}[hbt]
\epsscale{1.0}
\plotone{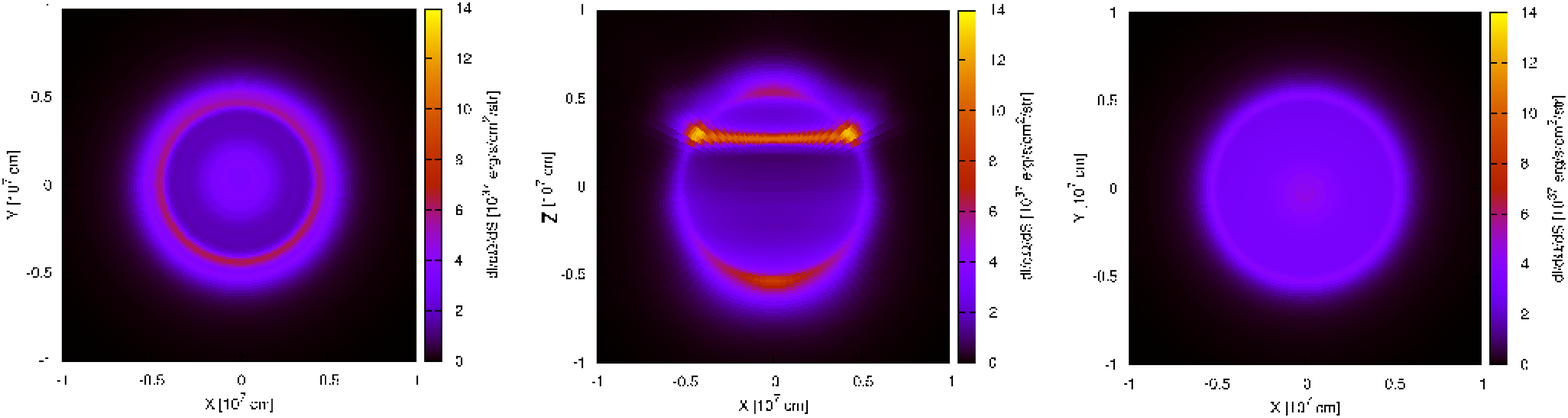}
\caption{Same as Figure \ref{f9} but for model B at $t=$330 ms, showing the 
 stronger emission seen from the equator (middle), as opposed to the cases of Figures 
\ref{f6} and \ref{f9}.}
\label{f11}
\end{figure}

\begin{figure}[hbt]
\epsscale{1.0}
\plotone{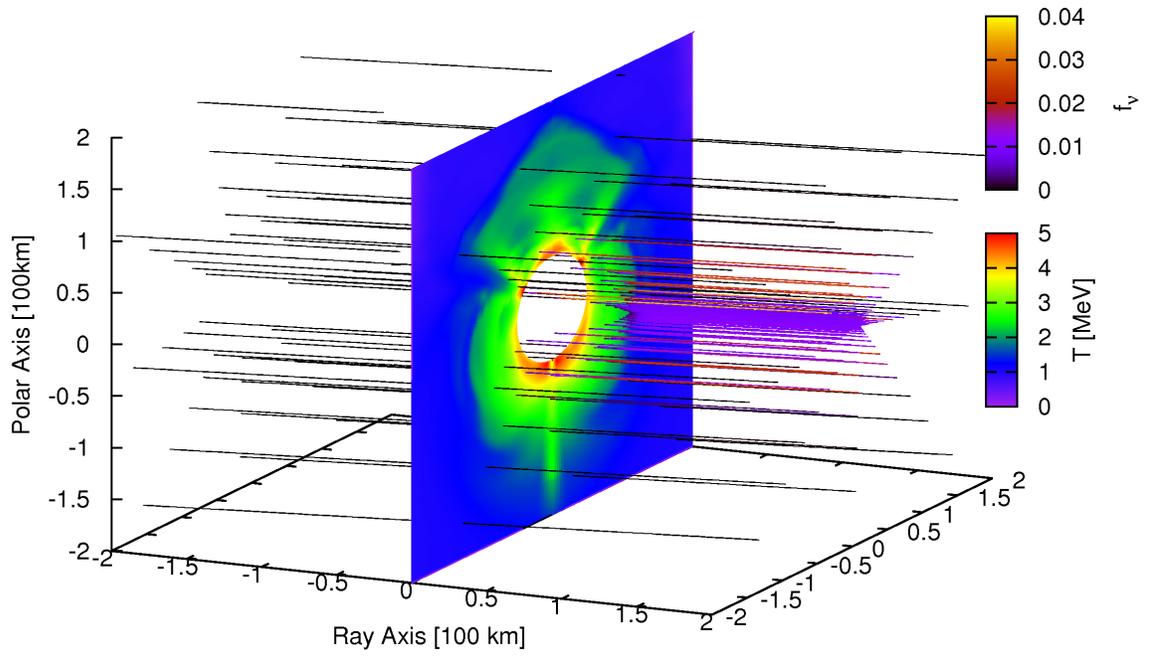}
\caption{Same as Figure \ref{f6} (middle right) but for model B 
 at $t=$330 ms.}
\label{f12}
\end{figure}

The middle panel of Figure \ref{f11} shows that the neutrino emission seen from 
the equatorial plane are much stronger than the ones seen from the northern/southern 
 hemispheres. The bright belt-like structure 
in the middle panel is from the high-temperature 
 regions in the northern hemisphere, which are by chance surrounded by relatively low 
density (see cusp-like structure in the northern hemisphere of Figure \ref{f10} and 
\ref{f12}). In this case, the area-weighted energy fluxes become larger seen from the 
equatorial plane (see right panel of Figure \ref{f8}). In combination with the 
 negative values of $\Phi(\theta^{'})$ in equation (\ref{tt}) in the vicinity of the 
 equatorial belts, this makes the negative growth in the GW amplitudes. Large negative 
 amplitudes seen for some other epochs in other model such as model C 
(left panel of Figure \ref{f3}) are also from the same reason.   
Such a feature is genuine outcome of the neutrino emission in the lateral direction, 
which is able to be captured correctly by the ray-tracing calculation.

It is noted that the appearance of 
the negative growth has no systematic dependence of the input luminosities.
 In fact, as seen from Figure \ref{f3}, the negative growth is observed 
for the intermediate luminosities models (models B and C), but not for the highest 
(model A) and smallest luminosity models (model D) (see also $|h_{\rm tot, max}|$ 
 in Table 1). 
 This should reflect the nature of the SASI which grows chaotically and non-locally.
 Albeit with the negative growth, our results suggest that the positively growing 
 features dominate over the negatively ones for the 2D models (see $h_{\nu, \rm{fin}}$ in  Table \ref{table1}). 
 This is due to the axial symmetry, along which the SASI develops preferentially and 
 the resulting anisotropies become larger. 

\begin{figure}[hbt]
\epsscale{1.0}
\plotone{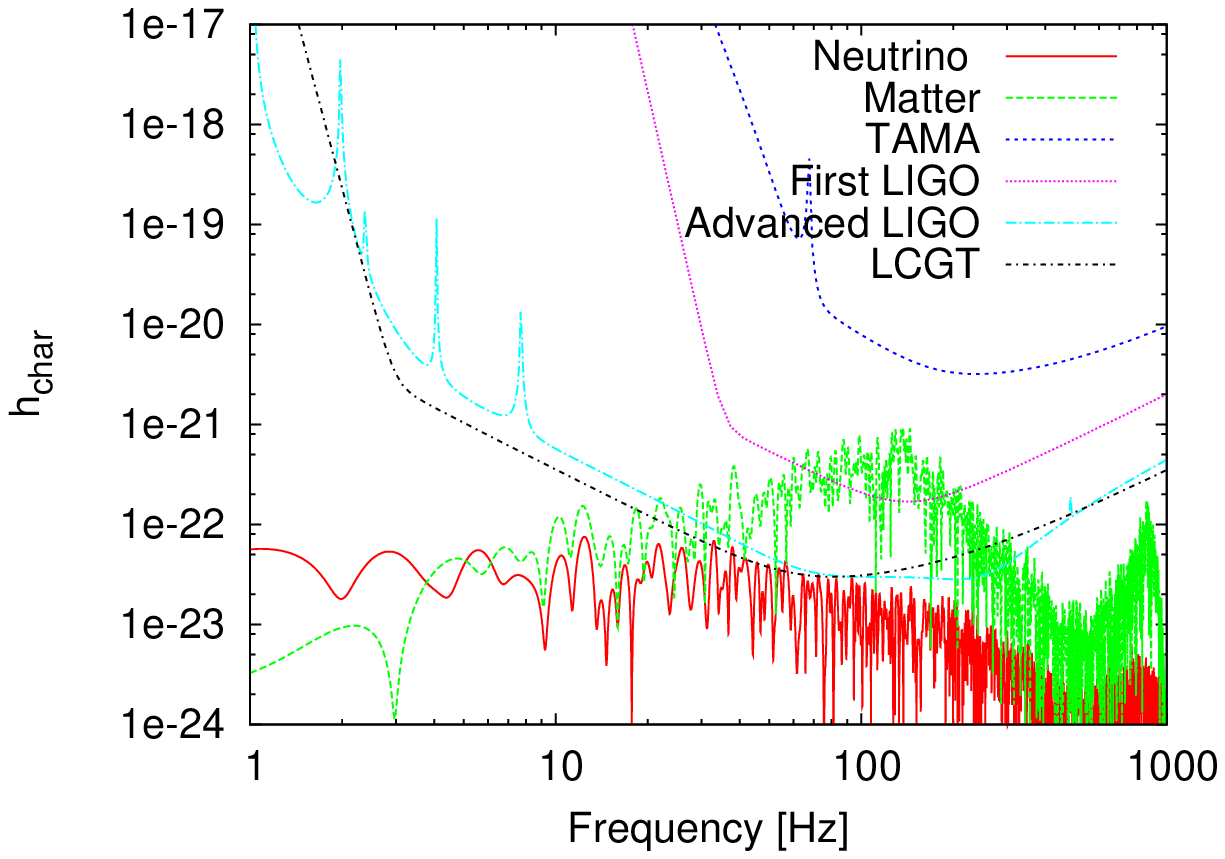}
\caption{Spectral distributions of GWs from matter motions (Matter) 
 and anisotropic neutrino emission (Neutrino) for model A with 
the expected detection limits of TAMA \citep{tama}, first LIGO \citep{firstligo}, 
advanced LIGO \citep{advancedligo}, and Large-scale
 Cryogenic Gravitational wave Telescope (LCGT) \citep{lcgt}.
 $h_{\rm char}$ is
 the characteristic gravitational wave strain defined in \citet{flanagan}. 
 Note that the supernova is assumed to be located at the distance of 10 kpc.}
\label{f13}
\end{figure}

As mentioned earlier, the neutrino GWs become 
more than one-order-of magnitude smaller than the previous estimation 
(compare $E_{{\rm GW,}~\nu}$ in Table \ref{table1} and the one in 
\citet{kotake_gw_sasi}).
 This stems not only from the incursion of the negative contributions but also from 
the appropriate estimation of the neutrino absorptions made possible 
by the ray-tracing method. 
Previously the neutrino 
luminosity was estimated simply by summing up the local neutrino cooling 
rates outside the PNSs \citep{kotake_gw_sasi},
 which fails to take into account the neutrino absorption correctly 
($\lambda$ in equation (\ref{transfer})). These two factors make the amplitudes much 
smaller than the previous estimation.
 As a result, the neutrino GWs, albeit dominant over the matter GWs 
in the lower frequencies below $\sim 10$ Hz (Figure \ref{f13}), 
 become very difficult to be detected for ground-based detectors whose sensitivity is 
limited mainly by the seismic noises at such lower frequencies 
\citep{tama,firstligo,advancedligo,lcgt}. 

On the other hand, the GWs from matter 
 motions seem marginally within the detection 
limits of the currently running detector of the first LIGO and the detection 
seems more feasible for the detectors in the next generation 
such as LCGT and the advanced LIGO for a Galactic supernova.
 The spectra of the matter GWs have double peaks namely near 
$100$ Hz and $1$ kHz. 
While the latter comes from the rapidly varying local 
hydrodynamical instabilities with milliseconds timescales,
the former is associated with the longer-term overturns of 
$O(10)$ ms induced by $\ell = 2$ mode of SASI (see e.g., 
Figure 5 in \citet{kotake_gw_sasi}). 
These gross properties in the GW spectra are common 
to the other luminosity models.
Thus the peak in the spectra near $\sim 100 $Hz 
 is found to be a characteristic feature obtained in the 2D models computed here.





\section{Summary and Discussion \label{sec5}}
We proposed a ray-tracing method to estimate gravitational waves (GWs) 
generated by anisotropic neutrino emission in supernova cores.
 To calculate the waveforms, we derived the GW formulae in a useful form, 
which are applicable also for 3D computations.
 Pushed by the studies supporting the slow rotation prior to core-collapse, 
 we considered an idealized situation that 
the neutrino radiation field from the protoneutron stars (PNSs) are isotropic. 
Then we focused on the asphericities outside the protoneutron stars,
 which are produced by the growth of standing 
 accretion shock instability (SASI).
 Since the regions outside the PNSs are 
basically thin to neutrinos,  we solve the transport equations by making use of 
the ray-tracing method in a post-processing manner.
For simplicity, neutrino absorption and emission 
 by free nucleons, dominant processes outside the PNSs, were only taken into account, 
 while the neutrino scattering and the velocity-dependent terms in the transport 
equations were neglected. 
Based on the two-dimensional models,
 which mimic SASI-aided neutrino heating explosions, 
 we estimated the 
 neutrino anisotropies by the ray-tracing method and calculated the resulting GWs. 


Our results show that the waveforms from neutrinos exhibit more variety 
in contrast to the ones previously estimated by ray-by-ray analysis (e.g.,
 \citet{kotake_gw_sasi}).
In addition to the positively growing feature, which was predicted to determine the 
 total wave amplitudes predominantly, the waveforms show
large negative growth for some epochs during the growth of SASI. 
 Such a feature is a genuine outcome of the neutrino emission in lateral directions, 
which can be captured correctly by the ray-tracing calculation.
Reflecting the nature of SASI which grows chaotically with time,
 little systematic dependence of the input neutrino luminosities on the 
maximum amplitudes and on the radiated GW energies are found.
 Due to the negative contributions and the neutrino absorptions appropriately 
taken into account by the ray-tracing method, 
 the GWs from neutrinos become more than one-order-of magnitude smaller 
than the previous estimation, making their detections very hard for a galactic source.
On the other hand, we point out that 
the gravitational-wave spectrum from matter motions have its peak 
 near  $\sim 100$ Hz,
  reflecting the growth of $\ell = 2$ mode of SASI with timescales of $O(10)$ ms.
Such a feature is found to be generic among 2D models computed here.

It should be noted that the approximations taken in the simulation, 
such as the excision inside the PNS with its fixed inner boundary and the light bulb
approach with the isentropic luminosity constant with time, 
are the very first step to model the dynamics of the neutrino-heating 
 explosion aided by SASI and study the resulting GWs.
 As already mentioned, the excision of the central regions inside PNSs 
 may hinder the efficient gravitational emission of the oscillating neutron star
\citep{ott_new}. It is recently reported that the waveforms obtained in the 
2D ray-by-ray Boltzmann simulations \citep{marek_gw} show a negatively growing feature. 
At first glance, our results may seem to contradict with theirs.
 They explained that the negative growth is due to the enhanced neutrino emission 
of muon and tau neutrinos near the equator of the PNS observed 
in their simulations.
 As we mentioned repeatedly, such feature inside the PNSs, cannot be captured 
 here in principle. Outside the PNSs, however, 
the stronger neutrino emission of $\nu_e$ and $\bar{\nu}_e$ 
along the polar regions are also seen in their simulations 
(see their bottom right panel of Figure 11 outside of $30 \sim
 40$ km, which is the surface of the PNS guessed from their Figure 1), 
which we can say, consistent with our results. 
As illuminated by this study, the elaborate estimation of the neutrino anisotropy
 is crucial to understand the gravitational radiation in supernova cores. 
 This naturally needs a full transport simulation
 coupled to 3D hydrodynamics encompassing the whole iron core, which is 
 beyond the scope of this study and a grand challenge 
for all the supernova modellers.

In addition, one more major deficit is the axial symmetry assumed in the 
present 2D simulations. In 3D, the pronounced dominance 
of $l =1,2$ along the symmetry axis, which is a coordinate singularity in the 
2D computations, may become weaker, owing to the additional spatial degree 
of freedom in the azimuthal direction. In the 3D case, 
we expect that the amplitudes become smaller owing to the reduced anisotropy 
along the symmetry axis \citep{iwakami08,iwakami08_2,iwakami_09}. Thus the 
amplitudes calculated in this study could be an upper bound, in which the
maximal anisotropy generated by SASI could be achieved. 
With the ray-tracing method and the GW formulae derived here,  we are now 
able to investigate the properties of
 the cross mode of GWs (e.g., \citet{kotake_3d}). We think it important because they 
 are of genuine 3D origin, which in combination of the plus modes, might give us some 
hints about the explosion asymmetry. Magnetic effects not only on the growth of 
 SASI \citep{endeve}, but also on the GW waveforms \citep{kotakegwmag} should be interesting.
This study is an appetizer before the forthcoming 3D studies to clarify
 those aspects, which will be presented elsewhere soon.

\acknowledgements{K.K. expresses thanks to K. Sato for continuing encouragements and 
 also to Yosuke Ogino for informative help.
We thank an anonymous referee
 for careful reading and a long list of useful comments.
Numerical computations were in part carried on XT4 and 
general common use computer system at the center for Computational Astrophysics, CfCA, the National Astronomical Observatory of Japan.  This
study was supported in part by the Grants-in-Aid for the Scientific Research 
from the Ministry of Education, Science and Culture of Japan (Nos. 19540309 and 20740150)
 and Grant-in-Aid for the 21st century COE program
``Holistic Research and Education Center for Physics of Self-organizing Systems''.}





\end{document}